# Renormalization and Knot Theory

Dirk Kreimer
Dept. of Physics
University of Tasmania
GPO Box 252C
Hobart
TAS 7001
Australia

December 8, 1994

**Abstract**

We investigate to what extent renormalization can be understood as an algebraic manipulation on concatenated one-loop integrals. We find that the resulting algebra indicates a useful connection to knot theory.

## 1 Introduction

Renormalization theory is one of the fundamental topics of perturbative Quantum Field Theory (pQFT). Renormalizability was one of the guiding principles in the construction of the Standard Model (SM), our current understanding of particle physics phenomenology. From first insights into the problem of ultraviolet (UV) divergences in a QFT [1] to the understanding of renormalization of gauge theories [2], including spontaneous symmetry breaking and the beautiful interplay with BRST identities [3], our understanding has vastly improved.

The first complete account to renormalization as such has been given by Dyson, Salam and Weinberg [4, 5, 6]. One can derive all properties of a renormalizable QFT from the knowledge of the Dyson Schwinger equations, an understanding of the problem of overlapping divergences [5] and Weinberg's theorem [6]. Later, the subject was reinvestigated thoroughly by [7]. Their approach was helpful especially in understanding parametric representations, but often the structure of Green functions is less transparent there. Nevertheless, the BPHZ method has become the standard text book approach to renormalization. Few would argue with the view that the structure of a renormalizable pQFT is most successfully encoded in structures like the forest formula [8], the BPHZ formalism and the $R$-operation [7].

Yet, as we will show, there seems to be still a more fundamental layer behind these structures. This paper is concerned with the exploration of an algebraic structure inherent in all renormalizable quantum field theories, which reveals a connection to knot theory and similar algebraic structures usually assigned to the investigation of braids. We will focus on this connection. We hope to show that these structures are not only of interest in their own right but also allow for a useful simplification of actual computation of renormalized quantities.

The paper is organized as follows. First, we will restrict ourselves to vertex corrections of ladder-type diagrams. This simple topology mainly serves to fix our notations and to exhibit the basic idea. Then we address our attention to the problem of overlapping divergences. In the further elaboration we follow the skeleton expansion as a guiding example. Thus we will show how to do the dressing of internal vertices and propagators in our approach, and thoroughly discuss the relation between the topology of a Feynman graph, the appearance of knots and the appearance of transcendental numbers in the MS $Z$ factors classifying the knots and thus the topology of the Feynman graph under consideration.

Our main objective is to show that, instead of calculating a Feynman diagram with all its associated counterterms, it is better to assign to it a corresponding link diagram (see below) and apply a skein relation to it (in the form of 'skein template algorithm' [9]) and conclude from the obtained link invariant polynomial the contribution to the $Z$-factor (in the MS scheme), generated by this diagram and its counterterms.

In fact, we will show that knot theory knows about the renormalization program in the sense that it generates all contributing forests with the help of the skein relation, revealing the proposed structure behind the recursion which governs renormalization theory.

We will restrict ourselves to MS schemes in the following but will comment on generalizations to other schemes in the course of our presentation. We will use dimensional regularization throughout the paper. We consider three-point couplings only, and will treat four-point couplings, as in $\lambda\varphi^4$ theory, elsewhere.

# 2 Ladder type vertex corrections

It is the purpose of this section to introduce the necessary notation to exhibit the basic idea underlying the proposed connection between knot theory and renormalization theory. We do so by investigating a simple ladder topology. This topology is one of the easiest accesible and serves as a convenient means to introduce our approach. Subsequent sections are devoted to the generalization to general graphs.

## 2.1 The Vertex Ladder

Consider a theory with renormalizable three-point couplings, which means for us a theory with vertex corrections of logarithmic degree of divergence. Let us restrict for a start to topologies of the following form:

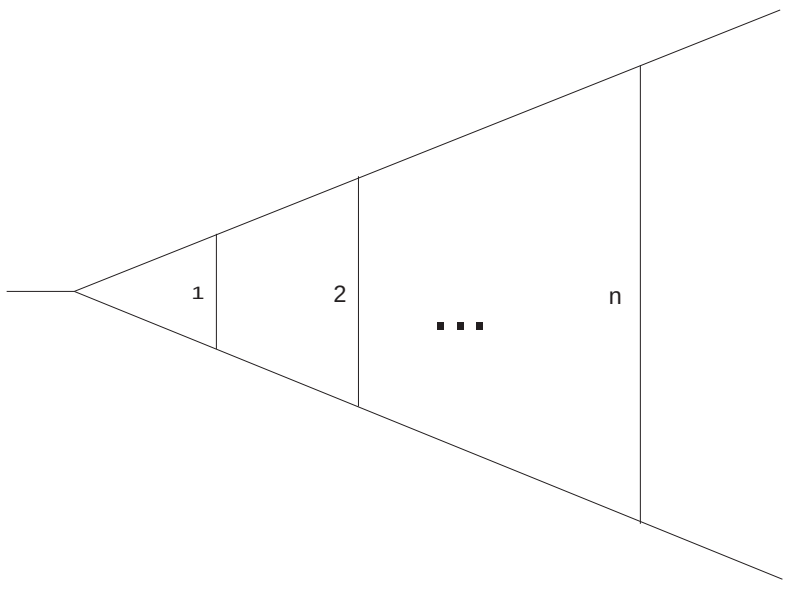

Fig.(1) The ladder topology for vertex corrections. We indicate loop momenta $l_1, \ldots, l_n$.

Note that the $n$-loop ladder involves $n-1$ rungs. The ladder topology might be realized by different Feynman graphs. Our graphical representation of Feynman diagrams usually refers only to the topology of the graph, so that lines may describe various different particles according to the Feynman rules of the theory. The above graphs all correspond to the

$$\Gamma \rightarrow \int \Gamma K \tag{1}$$

part of the Dyson-Schwinger equations. We have omitted all dressing of the internal vertices and propagators and pretend that there is only the ladder type vertex correction of the external vertex to be considered, thus we restrict ourselves to the first term in the skeleton expansion for the time being.

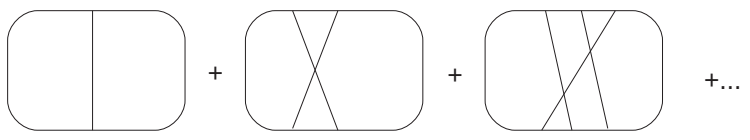

Fig.(2) The skeleton expansion.

For a start let us simplify even more and consider the two-loop case:

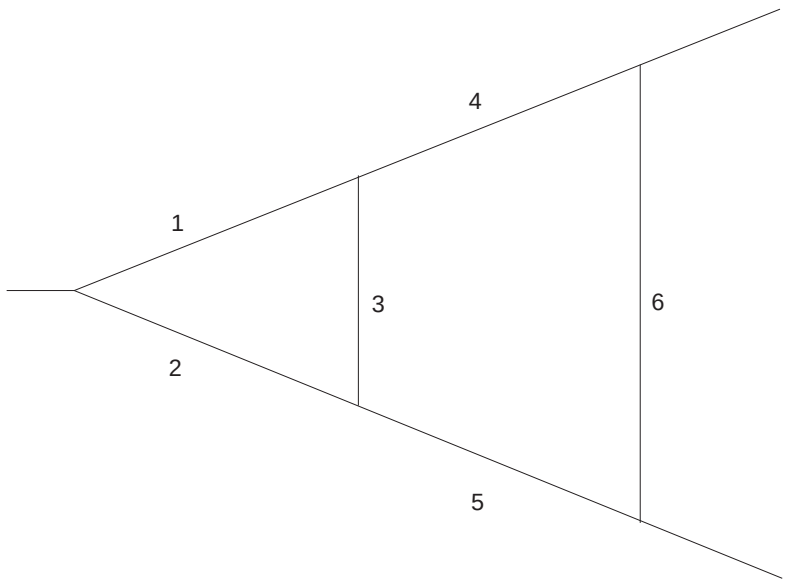

Fig.(3) Numbers correspond to possibly different masses in the propagators.

Normally, one would calculate this graph, then calculate its counterterm graphs (incorporating the one-loop $Z$-factor), and finally add these contributions. In the sum one would observe all the expected cancellations necessary to give a local two-loop $Z$-factor. This indicates that one might do better by subtracting out the subdivergences from the beginning. Thus, subtract from this vertex correction $\Gamma^{(2)}$ another vertex function, forming a new function $\Gamma^{(2)}_\Delta$:

$$\Gamma^{(2)}_\Delta := \Gamma^{(2)} - \tilde{\Gamma}^{(2)}, \tag{2}$$

where we define

$$\tilde{\Gamma}^{(2)} := \Gamma^{(2)} \mid_{m_1=m_2=m_3=p=0}, \tag{3}$$

that is in $\tilde{\Gamma}^{(2)}$ we have set all masses in the subdivergence to zero and evaluate the vertex at zero momentum transfer. $\Gamma^{(2)}_\Delta$ is a finite Green function:

$$\Gamma^{(2)}_\Delta = \int (\Gamma^{(1)} - \tilde{\Gamma}^{(1)})K,$$

is finite, as the integral kernel $K$ provides only a finite four-point function and $(\Gamma^{(2)} - \tilde{\Gamma}^{(2)})$ has a subtracted form.
Two examples:
For $\phi^3$ in six dimensions we have:

$$\tilde{\Gamma}^{(2)} = \int d^D l d^D k \, \frac{1}{l^4 (l+k)^2 (k^2 - m^2)^2 ((k+q)^2 - m^2)}, \tag{4}$$

and for massive QED in four dimensions, in the Feynman gauge:

$$\tilde{\Gamma}^{(2)}_\sigma = \int d^D l d^D k \, \gamma_\mu \frac{1}{\not k - m} \gamma_\rho \frac{1}{\not l} \gamma_\sigma \frac{1}{\not l} \gamma^\rho \frac{1}{\not k - m} \gamma^\mu \frac{1}{(l+k)^2} \frac{1}{(k+q)^2}, \tag{5}$$

corresponding to the Feynman graph in Fig.(4).

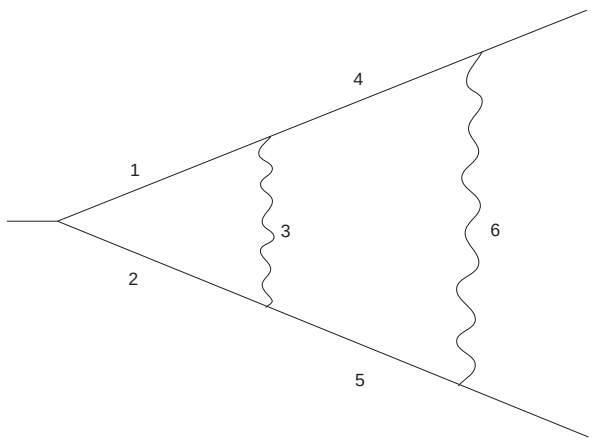

Fig.(4) The QED graph corresponding to the example Eq.(5).

We see that in both examples the $l$-integration over the subdivergence becomes particularly simple.

An important observation is that $\tilde{\Gamma}$ is not infrared divergent. It is a logarithmic UV-divergent vertex function evaluated at zero momentum transfer so that its infrared singularities are restricted to the on-shell domain which does not coincide with the set $\{0 = p = m_1 = m_2 = m_3\}$.

Now consider $\Gamma_\Delta$. It is UV convergent, both in its subdivergent and overall divergent behaviour, which can as well be concluded from power-counting. Our subtraction improves the power-counting for the inner loop momentum by one, but it improves also the overall degree of divergence by the same amount, as the subtracted term was only modified in exterior parameters like masses and momenta, and the overall degree of divergence is independent of these parameters. So far we have shown that the UV divergences of $\Gamma$ are located in a simpler Green function $\tilde{\Gamma}$, where we could set some masses to zero and evaluate at vanishing momentum transfer. From now on we make use of the following notation:

$$< \Gamma > = < \Gamma_\Delta + \tilde{\Gamma} > = < \Gamma_\Delta > + < \tilde{\Gamma} > = < \tilde{\Gamma} > . \qquad (6)$$

Here we used a projector $< \ldots >$ onto the UV-divergences (the proper singular part of a Laurent expansion of $\Gamma$ in $\varepsilon$, where $\varepsilon$ is the DR regularization parameter) so that we have

$$< \text{UV-finite expression} > = 0. \qquad (7)$$

We also use the $n$-th order complete $Z$-factor and various Green functions defined as follows:

$$\begin{aligned}
\mathbf{Z}_1^n &:= 1 - Z_1^{(1)} - Z_1^{(2)} \ldots - Z_1^{(n)}, \\
\mathbf{\Gamma}^n &:= \Gamma^{(0)} + \Gamma^{(1)} \ldots + \Gamma^{(n)}, \\
\tilde{\Gamma}^{(n)} &:= \Gamma^{(n)}|_{m_1=m_2=m_3=p=0}, \\
\bar{\Gamma}^{(n)} &:= \tilde{\Gamma}|_{m_i=0\forall i}.
\end{aligned} \qquad (8)$$

Here, the index $n$ refers to the loop order. Now, having absorbed the UV-divergence of $\Gamma^{(2)}$ in $\tilde{\Gamma}^{(2)}$ let us add the counterterm graph

$$\tilde{\Gamma}^{(2)} \to Z_1^{(2)} := < \tilde{\Gamma}^{(2)} - Z_1^{(1)} \tilde{\Gamma}^{(1)} >, \qquad (9)$$

$Z_1^{(2)}$ contains $<\Gamma^{(2)}>$ plus its counterterm graph. We will become more accustomed with its use in the following.

With the above definitions we have

$$<\tilde{\Gamma}^{(2)} - \bar{\Gamma}^{(2)} - Z^{(1)}(\tilde{\Gamma}^{(1)} - \bar{\Gamma}^{(1)}) >= 0, \tag{10}$$

since the overall divergent behaviour of

$$(\tilde{\Gamma}^{(2)} - Z_1^{(1)}\tilde{\Gamma}^{(1)}),$$

and

$$(\bar{\Gamma}^{(2)} - Z_1^{(1)}\bar{\Gamma}^{(1)}),$$

are the same. This is exactly what we expect. Both expressions have the same asymptotic behaviour with regard to their overall degree of divergence, and have the same subdivergence $Z_1^{(1)}$. Necessarily, the difference is UV convergent. This is nothing else than the statement that $Z_1^{(2)}$ must be polynomial in exterior masses and momenta.

By use of Eq.(10) we have absorbed all UV singularities in the expression

$$Z_1^{(2)} = \bar{\Gamma}^{(2)} - Z_1^{(1)}\bar{\Gamma}^{(1)}. \tag{11}$$

Only massless three-point functions at zero momentum transfer appear in the above expression.

Now, the one-loop $Z$-factor from the vertex correction, $Z_1^{(1)}$, has the form (with momentum $r$ say flowing through the diagram)

$$\bar{\Gamma}^{(1)} = \Delta^1(\varepsilon)\ (r^2)^{-\varepsilon}\Gamma^{(0)} \Rightarrow Z_1^{(1)} = \bar{\Gamma}^{(1)}(r^2)^{\varepsilon} - c_0, \tag{12}$$

where we used the fact that $\bar{\Gamma}^{(1)}$ scales like $(r^2)^{-\varepsilon}$ in DR and we assume that the usual $\mu$-scaling is absorbed in a redefinition of the coupling constant; $\Gamma^{(0)}$ is the tree-level vertex. As we are calculating in a MS scheme, for the present we have subtracted out the finite part $c_0$ in the Laurent expansion of $\bar{\Gamma}^{(1)}(r^2)^{\varepsilon}$.

A vital aspect of this approach is that we can always write $\bar{\Gamma}^{(1)}$ as

$$\int d^Dk\ \ F(k,q) \equiv \bar{\Gamma}^{(1)} = \Delta^1\ (q^2)^{-\varepsilon}\Gamma^{(0)}, \tag{13}$$

where $\Delta^1$ is some function of $\varepsilon$. With these definitions we have

$$Z_1^{(1)} =< \Delta^1 >, \tag{14}$$

the bracket doing the job of the MS-operation.

Let $F(k,r)$ be the integrand for the calculation of $\bar{\Gamma}^{(1)}$, cf. Eq.(13). We see that (using Eq.(13))

$$\bar{\Gamma}^{(2)} = \Delta^1 \int d^Dk (k^2)^{-\varepsilon} F(k,q) = \Delta^1\ {}_1\Delta^1\ \ (q^2)^{-2\varepsilon}\ \Gamma^{(0)}, \tag{15}$$

where we define for later use ${}_j\Delta^1$ by

$$\int d^Dk(k^2)^{-\varepsilon j}F(k,q) =: {}_j\Delta^1 \;\; (q^2)^{-\varepsilon(j+1)}\,\Gamma^{(0)},$$
$$(\Delta^1 \equiv {}_0\Delta^1). \tag{16}$$

We consider the functions ${}_j\Delta^1$ as modified one-loop functions. For any renormalized theory they can be obtained from the corresponding standard one-loop integral by a change in the measure

$$\int d^Dk \rightarrow \int d^Dk(k^2)^{-\varepsilon j}.$$

Here we tacitly assumed that the massless one-loop vertex function at zero momentum transfer only reproduces the tree-level vertex $\Gamma^{(0)}$ as a form factor (Eq.(13)). This is not true for all three-point vertices appearing in the SM, for example. The exceptional case there is the fermion coupling to the gauge boson. It turns out that the modifications due to this complication do not spoil our general reasoning. We will present the necessary changes in the appendix.

We denote the parameter $j$ above as the 'writhe number' in the following, for reasons which become clear below.

We finally have in our two-loop example

$$Z_1^{(2)} =< \Delta^1 \; {}_1\Delta^1 - < \Delta^1 > \Delta^1 >, \tag{17}$$

by virtue of Eq.(13,14,15).

Our next step will be to generalize the above approach to the $n$-loop case. We still pretend that the world consists only of ladder-type vertex corrections, in which case our approach would deliver the full $Z$-factor for the vertex correction. The general case is the subject of later sections.

Again we define

$$\Gamma_\Delta^{(n)} := \Gamma^{(n)} - \tilde{\Gamma}^{(n)}, \tag{18}$$

always evaluating at zero momentum transfer and setting the masses $m_1, m_2, m_3$ of the inner loop to zero. It follows that $< \Gamma_\Delta^{(n)} >= 0$. This is obvious as all existing forests include the innermost loop as a nest. All divergent sectors (which are all logarithmic divergent) thus have an improvement in power counting when we have such an improvement for this inner loop. It follows also by considering $\Gamma_\Delta = \int \ldots \int (\Gamma^{(1)} - \tilde{\Gamma}^{(1)})K \ldots K$, and writing out the iteration of the bubble in the Dyson-Schwinger equations explicitly.

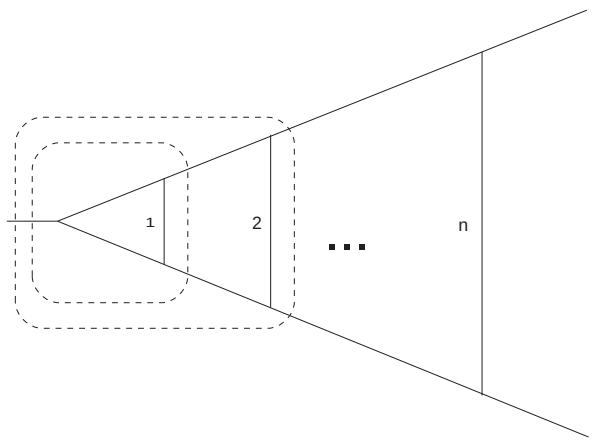

Fig.(5) The brackets denote the divergent sectors of the ladder topology. These give rise to the restricted set of counterterms we are interested in. The divergent sectors are, from the left to the right, the domains $\lambda \rightarrow \infty$ in $\{\lambda l_1\}, \{\lambda l_1, \lambda l_2\}, \ldots, \{\lambda l_1, \ldots, \lambda l_{n-1}\}$.

To calculate $Z_1^{(n)}$ we have to consider

$$Z_1^{(n)} = < \bar{\Gamma}^{(n)} - \sum_{i=1}^{n-1} Z_1^{(i)} \bar{\Gamma}^{(n-i)} >, \tag{19}$$

In this expression we are allowed to consider massless quantities only as the overall degree of divergence is mass independent, and the term in brackets is free of subdivergences by construction.

We have

$$\bar{\Gamma}^{(n)} = (q^2)^{-\varepsilon n} \prod_{i=0}^{n-1} {}_i\Delta^1, \tag{20}$$

according to our definition Eq.(15). Here sequential expressions like $\Delta^1{}_1\Delta^1 \ldots$ denote a concatenation of one-loop functions.

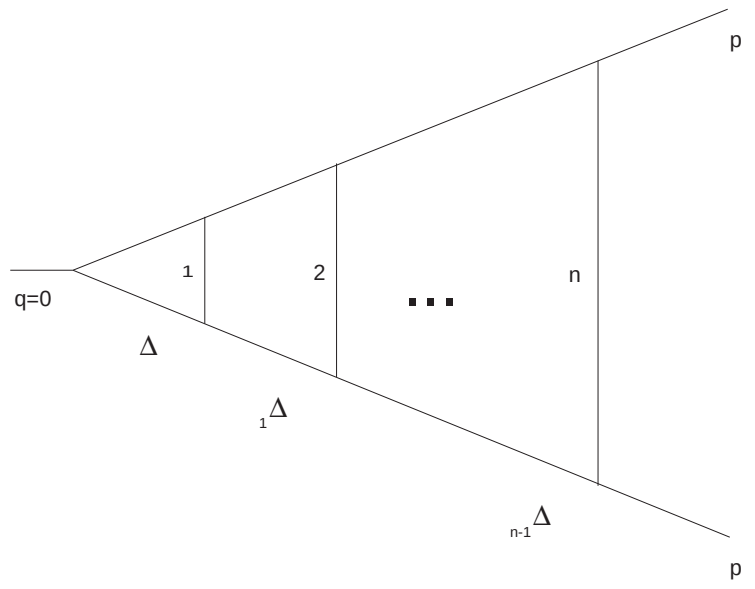

Fig.(6) By evaluating at zero momentum transfer the calculation of a massless ladder becomes a concatenation of ${}_j\Delta^1$ functions.

Again, in case that there should be more than one form factor in this concatenation some technical subtleties are involved which are given in the appendix in detail.

We define two operators

$$\begin{aligned} B^k(\Delta^1) &:= \prod_{i=0}^{k} {}_i\Delta^1, \\ A^r(\alpha) &:= \Delta^1 <<< \ldots < \alpha > \Delta^1 > \ldots \Delta^1 >, \\ &\quad \text{r <> brackets} \\ \Rightarrow A^r(B^k(\Delta^1)) &= \Delta^1 <<<< \ldots < \prod_{i=0}^{k} {}_i\Delta^1 > \Delta^1 > \ldots \Delta^1 >, \\ B^k(A^r(\Delta^1)) &= <<< \ldots < \Delta^1 > \Delta^1 > \ldots \Delta^1 > \prod_{i=0}^{k} {}_i\Delta^1, \\ B^0(\Delta^1) = A^0(\Delta^1) &= \Delta^1. \end{aligned} \tag{21}$$

$B$ acts by concatenating massless one-loop functions with increasing writhe number, $A$ by projecting on the divergent part of products iteratively.

We can now give the general result for $Z_1^{(n)}$:

$$Z_1^{(n)} =< [-A + B]^{(n-1)}(\Delta^1) > . \tag{22}$$

For $n = 3$ this gives explicitly

$$\begin{aligned} Z_1^{(3)} =< \Delta^1 \,{}_1\Delta^1 \,{}_2\Delta^1 - (< \Delta^1 \,{}_1\Delta^1 > - << \Delta^1 > \Delta^1 >)\Delta^1 \\ - < \Delta^1 > \Delta^1 \,{}_1\Delta^1 > . \end{aligned} \tag{23}$$

We remind the reader that the functions ${}_j\Delta^1$ are simple modifications of the corresponding three-point one-loop function, evaluated in the massless limit at zero momentum transfer. The above concatenation properties are universal for all these functions, as long as they are generated by a renormalizable theory. Note that there might be in fact a whole 'bouquet' of these functions at our disposal, as in more realistic theories our ladder topology might consist of various different $\Delta$'s, as in the example of Fig.(7):

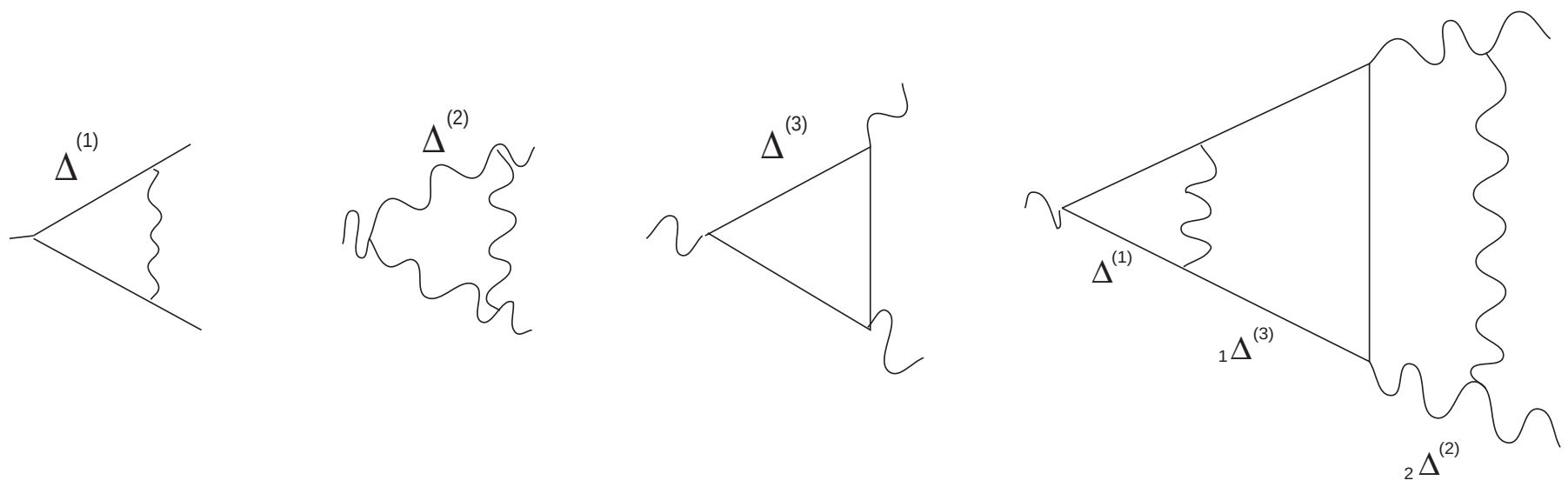

Fig.(7) Various different $\Delta$'s. We understand that in our result Eq.(22) we have to consider the appropriate combination of $\Delta$'s as arguments for the $A$ and $B$ operators.

Recalling our definition for the full vertex function

$$\begin{aligned}\mathbf{\Gamma}^{(n)} &:= \Gamma^{(0)} + \sum_{i=1}^{n} \Gamma^{(i)}, \\ \Rightarrow < \mathbf{\Gamma}^n > &= < \bar{\mathbf{\Gamma}}^n >, \end{aligned} \quad (24)$$

we have its MS renormalized form,

$$\mathbf{\Gamma}_R^n = \mathbf{Z}_1^n \mathbf{\Gamma}^n. \quad (25)$$

We can easily check that $\mathbf{Z}_1^{(n)}$ renders $\mathbf{\Gamma}_R^n$ finite. Indeed, we only have to check that $\bar{\mathbf{\Gamma}}_R^n$ is finite and easily calculate this expression to be (using Eq.(19,22))

$$\begin{aligned}\bar{\mathbf{\Gamma}}_R^n &= (1 + \sum_{i=0}^{n} [B]^i(\Delta^1))(1 - \sum_{i=0}^{n-1} < [-A+B]^i(\Delta^1) >), \\ &= 1 + \sum_{i=0}^{n-1} ([-A+B]^i(\Delta^1) - < [-A+B]^i(\Delta^1) >), \end{aligned} \quad (26)$$

which is evidently finite. In some obvious shorthand notation we can write this result to all orders as

$$\bar{\mathbf{\Gamma}}_R = 1 + \frac{1}{1+[-A+B]}(\Delta^1) - < \frac{1}{1+[-A+B]}(\Delta^1) > . \quad (27)$$

Here, an expansion in the coupling constant is understood ($A^r B^m$ is of order $\mathcal{O}(g^{2(r+m)})$). Before we extend this rather academic example to more realistic situations, including the other topologies, self energies and the like let us summarize what we have achieved so far. We have calculated the ladder-type vertex corrections of Fig.(1) above, including the counterterms as depicted in Fig.(8):

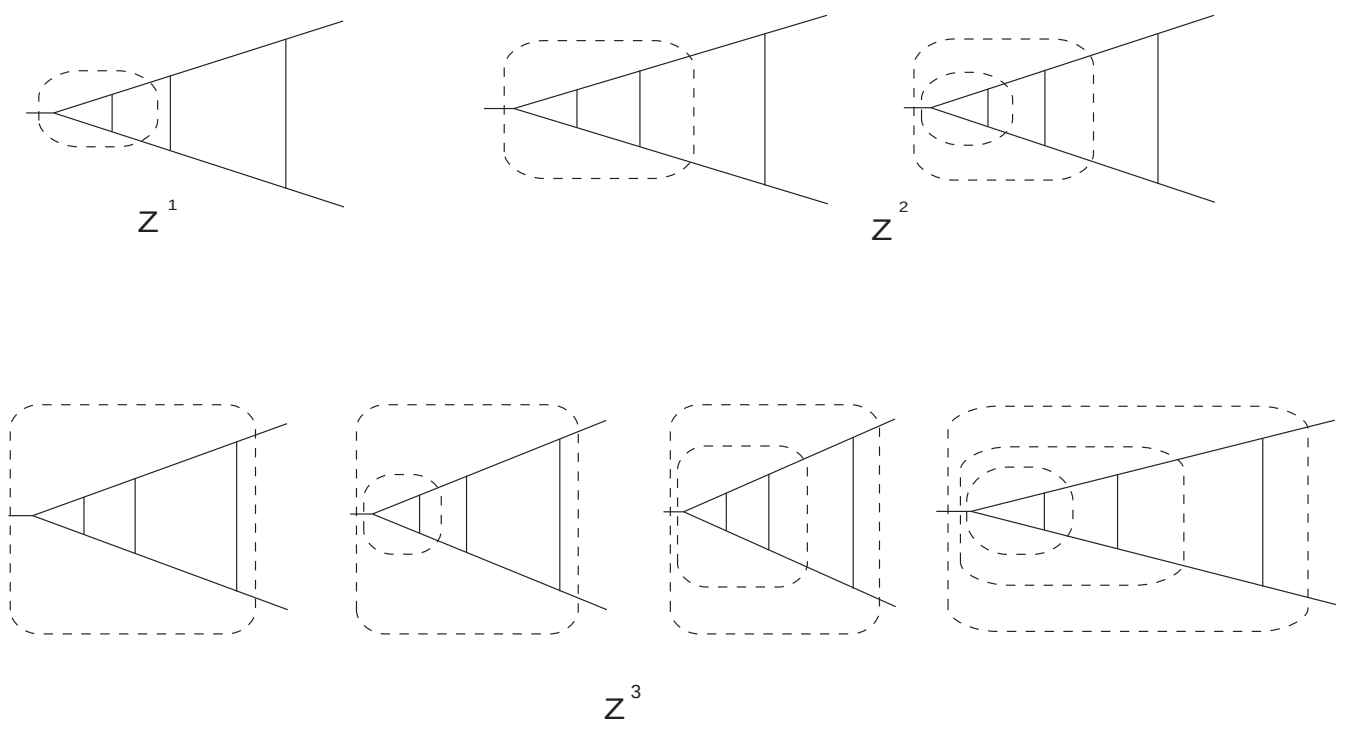

Fig.(8) We have calculated $< \Gamma^{(n)} >$ so far, including the counterterms for the dashed divergent sectors, taking into account all possible forests.

Our main objective has to be to generalize this to all possible topologies, including dressing of internal lines and vertices, and, crucially, including overlapping divergences. But before we do so in the next sections, let us have another look at our ladder. At this stage we have introduced a sufficient amount of notation so that we can start to consider our example again, this time turning to knot theory as the approach to obtain our results Eq.(22).

## 2.2 Links and Ladders

In the following we consider link diagrams. We use some elementary notions of link and knot theory, as provided for example in [9]. We regard link diagrams consisting of $n$ components say, that is $n$ rings tangled in each other in various ways. According to the rules we give below, the single components in the link diagrams of interest are all trivial circles (unknots). Only through the entanglement we generate non-trivial topological structure in the link diagrams. Using skein relation to disentangle the diagrams we will then generate knots in intermediate steps of this algorithm, in fact applying the skein relation $n$ times to a $n$ component link will generate one-component knots amongst other terms. We claim that these knots classify our Feynman graph and determine the UV divergences.

For a start, let us introduce the following two rules to map any Feynman diagram into a link diagram.

- Every loop in the Feynman diagram corresponds to a link. Correspondingly, a $n$-loop diagram will map to a link diagram consisting of $n$ links.
- The links are oriented according to the flow of loop momenta, and follow the rule that at every vertex the momentum coming from the right is overcrossing:

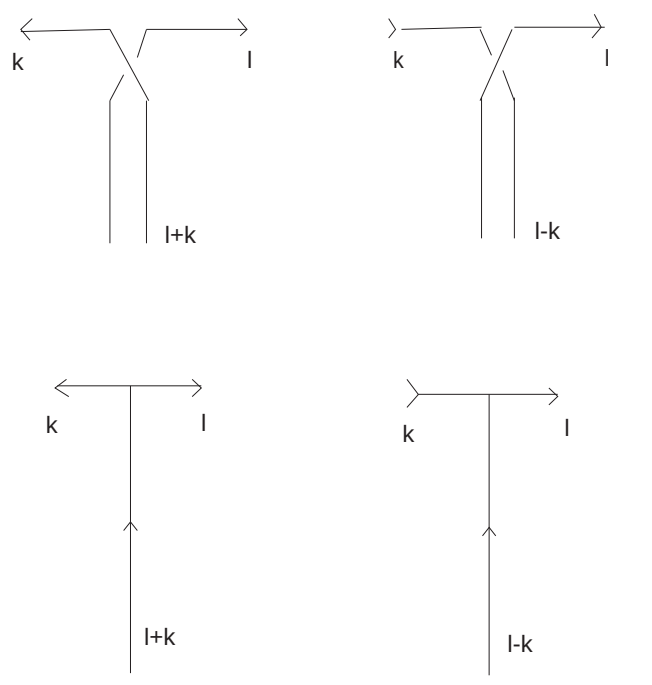

Fig.(9) The replacement of a three-point vertex by an overcrossing.

Ignore exterior momenta in this process (for the moment).

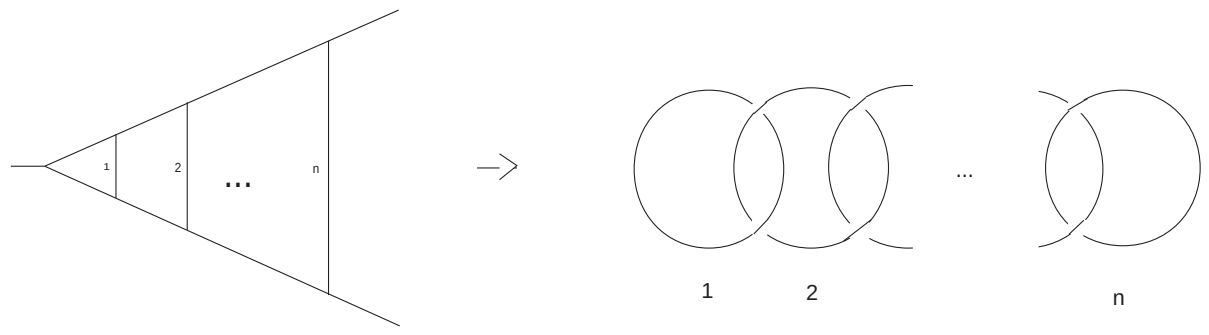

Fig.(10) The translation from a Feynman diagram into a link diagram. Each vertex is replaced by an over/undercrossing according to the momentum flow at the vertex. We follow the convention to have the momentum flow in each loop counterclockwise.

Typically, our nested loop structure gives us a sequence of (Hopf-) links.

We now presume that some sort of braid structure underlies the algebraic systematics of renormalization theory. So we assume that we can establish a skein relation of the form

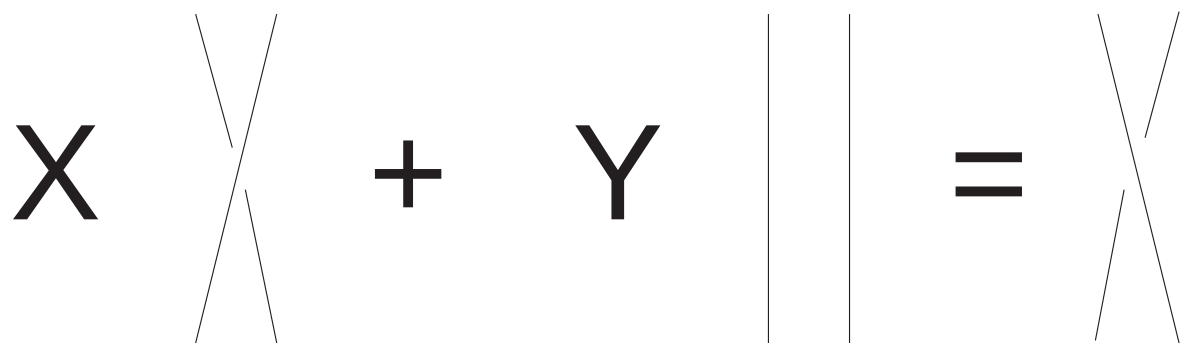

Fig.(11) The skein relation, an exchange identity which allows the disentangling of the link diagram. $X$ and $Y$ have to be regarded as operators to be identified with $A$ and $B$ in an appropriate manner.

This assumption is at the moment based only on the vague evidence that the ladder topologies renormalize according to Eq.(22), which, as we will see below, fits into the pattern suggested by a skein relation. In the course of the following sections we will look for further evidence to justify this assumption.

Let us consider the two-link diagram and its disentanglement.

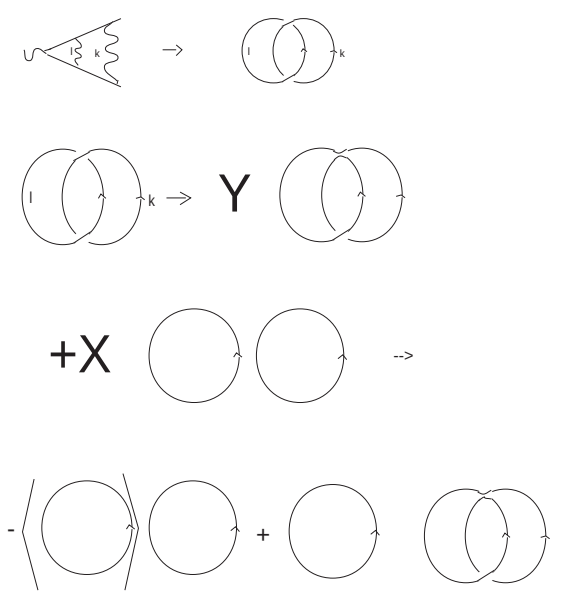

Fig.(12) A two-loop example explicitly.

For the three-link diagram, corresponding to $\Gamma^{(3)}$, we have

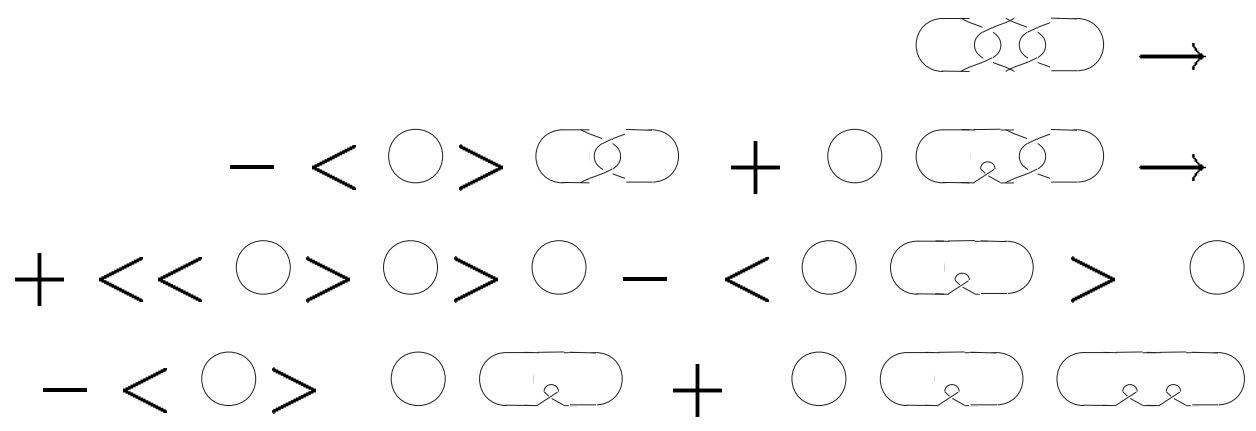

Fig.(13) The same argument for the three-loop case.

With the identification

$$\begin{aligned} X^{r-1}(\bigcirc\ \bigcirc \ldots \bigcirc\ ) &\Rightarrow [-A]^{k-1}(\Delta^1), \\ Y^{r-1}(\ \ldots\ )_{r-1} &\Rightarrow B^{r-1}(\Delta^1), \end{aligned} \tag{28}$$

we obtain our previous result. We identify the unknot with an appropriate one-loop function $\Delta^1$ and links of the form with the corresponding function ${}_j\Delta^1$. Note that this implies that we have no invariance under Reidemeister type $I$ moves, so that we work with a regular isotopy.

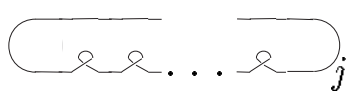

Fig.(14) A link with writhe number $j$.

We conclude that the $n$-link diagram of the form of Fig.(10) delivers, via the skein relation and appropriate identification of the operators $A$ and $B$, the $Z$-factor, $Z_1^{(n)}$.

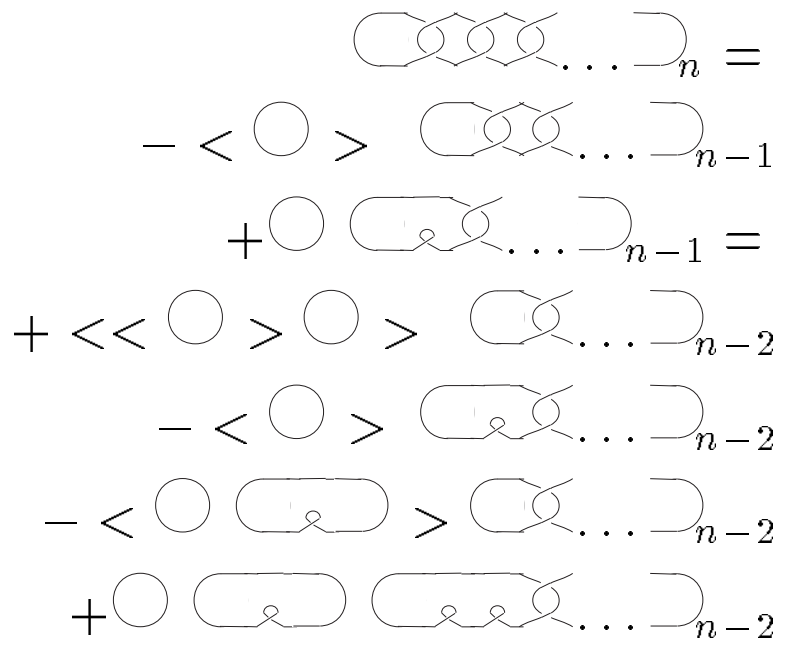

Fig.(15) The skein tree in general.

Let us briefly discuss how the above approach looks from the point of view of a braid group approach. Every closed link entangled in some other closed link generates a braid diagram with a very peculiar topology:

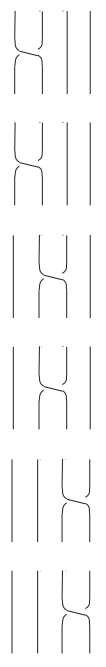

$= \sigma_1^2 \sigma_2^2 \sigma_3^2$

Fig.(16) The braid diagram. A closure of all strands is always understood. The above braid would be generated in the case $n = 4$.

Every crossing of a line $i$ with a line $(i{+}1)$ as shown above corresponds to the action of a braid group generator $\sigma_i$. Note that all braids resulting from Feynman diagrams can be described by positive powers of the braid group generators, according to our overcrossing rule.

The $n$-link diagram Fig.(10) we identify with $Z_1^{(n)}$

$$\begin{aligned} \sigma_1^2 \ldots \sigma_{n-1}^2 &\leftrightarrow Z_1^{(n)}, \\ \sigma_i^2 &= Y\sigma_i + X\mathbf{1}, \end{aligned} \tag{29}$$

which implies the usual identification of the skein relation with the Hecke algebra relation $\sigma = X\sigma^{-1} + Y\mathbf{1}$, so that we would recover our previous expressions, e.g.

$$\begin{aligned} Z_1^{(3)} &\leftrightarrow \sigma_1^2\sigma_2^2 \Rightarrow \\ &= X^2 - XY\sigma_1 - YX\sigma_2 + Y^2\sigma_1\sigma_2 \Rightarrow \\ &= [-A+B]^2(\Delta^1). \end{aligned} \tag{30}$$

Here we used a Hecke algebra representation of the braid group. At this point the algebra is not very specific. Our $n$-link corresponds to words containing nothing else than products of the form $\prod_i \sigma_i^2$, so that we can draw the braid diagram in the simple block form of Fig.(16). Due to this simple structure we do not, at this level, see much of the proper braid algebra structure, encoded in relations

$$\sigma_i\sigma_{i+1}\sigma_i = \sigma_{i+1}\sigma_i\sigma_{i+1}. \tag{31}$$

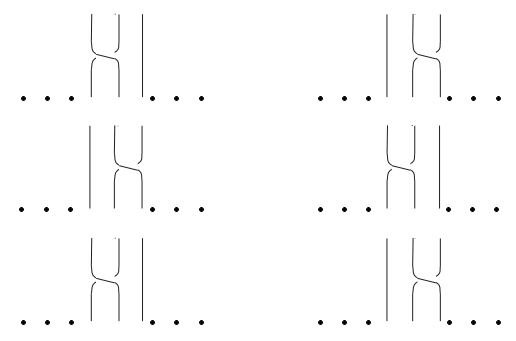

Fig.(17) The Reidemeister type $III$ move corresponding to Eq.(31). It does not appear in the $n$-link diagrams studied so far. The type $II$ move , corresponding to $\sigma_i\sigma_i^{-1} = 1$, was sufficient so far.

We will not dwell on this point here but will have more to say later on.

## 2.3 A Generalization

This generalization shows mainly that also more complicated ladder type graphs have a behaviour which fits into the pattern established so far. Nevertheless they correspond to more complicated link diagrams not encountered yet. For the time being we regard any $n(\alpha_i)$-loop topology $\alpha_i$ below as corresponding to a cable (this cable being a $n(\alpha_i)$ component link itself) and ignore the 'fine structure' of these cables.

So let us now slightly generalize the ladder topology of Fig.(6) above. We want to include all sorts of vertex corrections which are themselves free of subdivergences. This corresponds to a full skeleton expansion of the vertex:

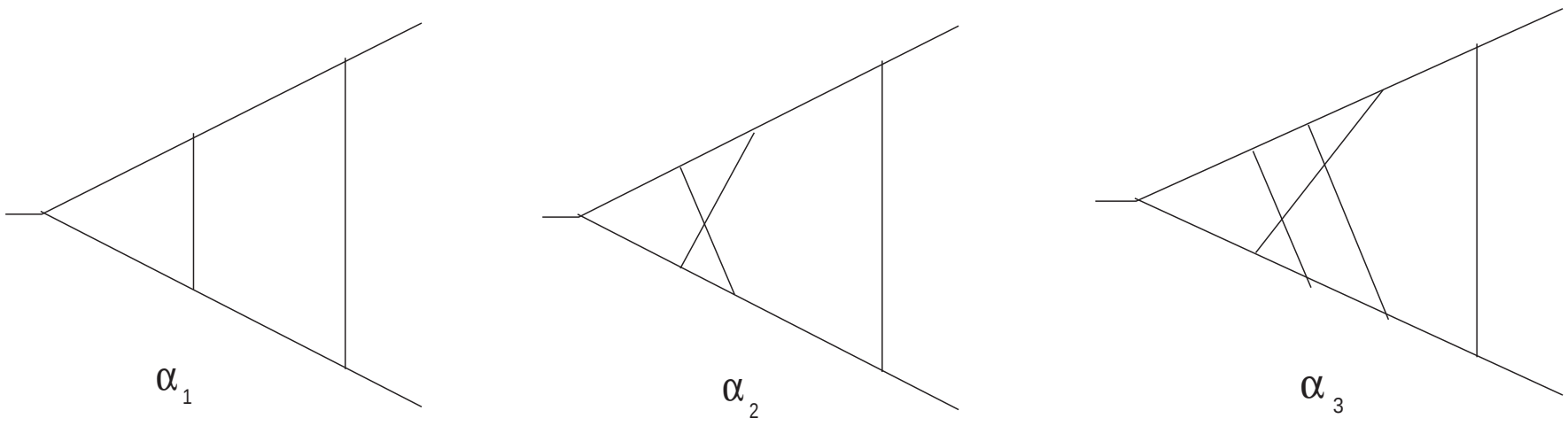

Fig.(18) The general ladder topology. Every cable of lines defining a new subdivergence is labelled according to its topology.

We label the different topologies by indices $\alpha_i$ but still omit the dressing of internal vertices and propagators. Note that every topology gives rise to a simple pole in $\varepsilon$ only

$$\Delta_{\alpha_i} = \frac{c_{\alpha_i}}{\varepsilon} + d_{\alpha_i}, \tag{32}$$

due to the fact that it has no subdivergences. Let $n(\alpha_i)$ be the number of loops in $\Delta_{\alpha_i}$. A Feynman graph of the form

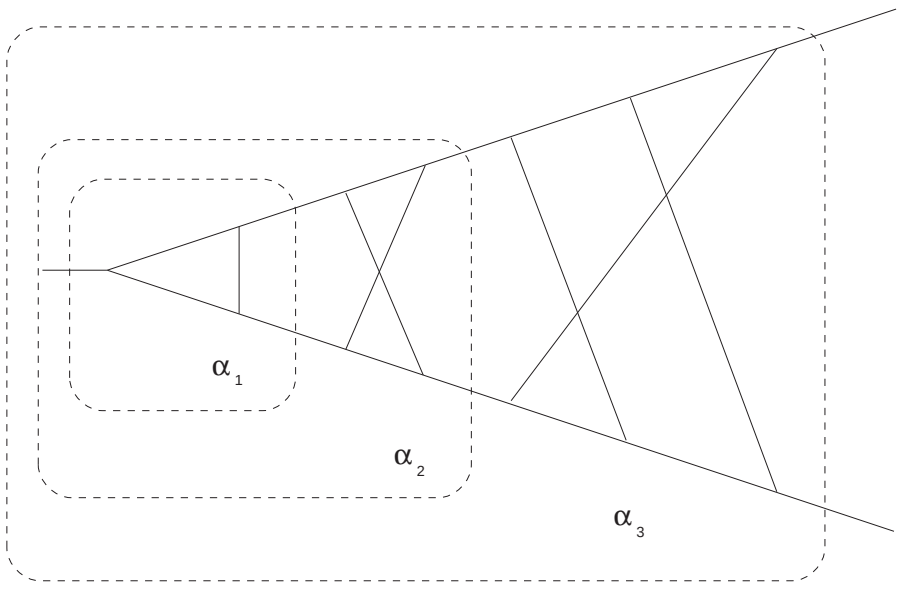

Fig.(19) The cabling of loops into divergent sectors.

will then result in a $Z$-factor contribution

$$Z \;=\; \prod_{i=1}^{n-1}[-A+B](\Delta_{\alpha_i}), \tag{33}$$

with the obvious definitions

$$\begin{aligned}
\prod_{i=r}^{s} A(\Delta_{\alpha_i}) &:= \Delta_{\alpha_r} <<<< \Delta_{\alpha_{r+1}} > \Delta_{\alpha_{r+2}} > \ldots \Delta_{\alpha_s} >, \\
\prod_{i=r}^{s} B(\Delta_{\alpha_i}) &:= \Delta_{\alpha_r}\; {}_{n(\alpha_r)}\Delta_{\alpha_{r+1}} \cdots {}_{n(\alpha_r)+\ldots+n(\alpha_{s-1})}\Delta_{\alpha_s}.
\end{aligned} \tag{34}$$

Note that the concatenation in the $B$ operator takes the loop number of each topology into account.

From the viewpoint of braids we would still map this situation to a $n$-cable link, so that we associate a cable (a collection of links) to each $\Delta^{\alpha_i}$. In braid generator language we would obtain generators $\sigma$, each acting on links in a different representation $\alpha_i$ so to speak, where in the '$B$' part of the skein relation we have a concatenation of writhe numbers (framing numbers) as shown above (taking the various $n(\alpha_i)$'s into account). An example of this is Fig.(19), which delivers:

$$\begin{aligned} Z_1^{(3)}(\alpha_1,\alpha_2,\alpha_3) =& < \Delta^1_{\alpha_1\ n(\alpha_1)} \Delta^1_{\alpha_2\ n(\alpha_1)+n(\alpha_2)} \Delta^1_{\alpha_3} \\ & -(< \Delta^1_{\alpha_1\ n(\alpha_1)} \Delta^1_{\alpha_2} > - << \Delta^1_{\alpha_1} > \Delta^1_{\alpha_2} >) \Delta^1_{\alpha_3} \\ & - < \Delta^1_{\alpha_1} > \Delta^1_{\alpha_2\ n(\alpha_2)} \Delta^1_{\alpha_3} > . \end{aligned} \tag{35}$$

The $\Delta^1_{\alpha_i}$ are implicitly defined as the dimensionless function of the regularization parameter $\varepsilon$ which multiplies the scaling $(r^2)^{-\varepsilon n(\alpha_i)}$ of the corresponding $n(\alpha_i)$-loop Green function.

Note that our presentation here was done under the assumption that the $\Delta_{\alpha_i}$ are known. This is true for all possible topologies up to the four-loop level, thanks to the major progress in massless two-point functions, obtained by various authors [11]. A summary of the situation in this area can be found in [12]. Later we will see that our knot-theoretic approach suggests a way to obtain these functions to all loop orders for all topologies by associating them with various knots.

To get the final $Z$-factor we just have to add the results for the various topologies. Define the total loop order $n_t$ by

$$n_t = \sum_i n(\alpha_i). \tag{36}$$

We have for the $Z$-factor generated from all topologies contributing in a given loop order $m$

$$\mathbf{Z}_1^n = 1 - \sum_{m=1}^{n} \sum_{I_m} Z_1^{(n_t=m)}(I_m), \tag{37}$$

where $\{I_m\}$ denotes a complete set of topologies such that $n_t = m$, and the sum over $I_m$ runs over all these topologies. Note also that with different topologies the number of different types of graphs per topology proliferates and has to be taken into account by an appropriate choice of basic functions $\Delta_{\alpha_i}$. This ends our considerations of the vertex ladder topology and we now turn to two-point functions.

# 3 Rainbow type self-energy corrections

Again we consider only nested divergences. The whole situation is largely a repetition of the situation in the previous section. Nevertheless, we have to comment on some new features. Consider the following general rainbow graph.

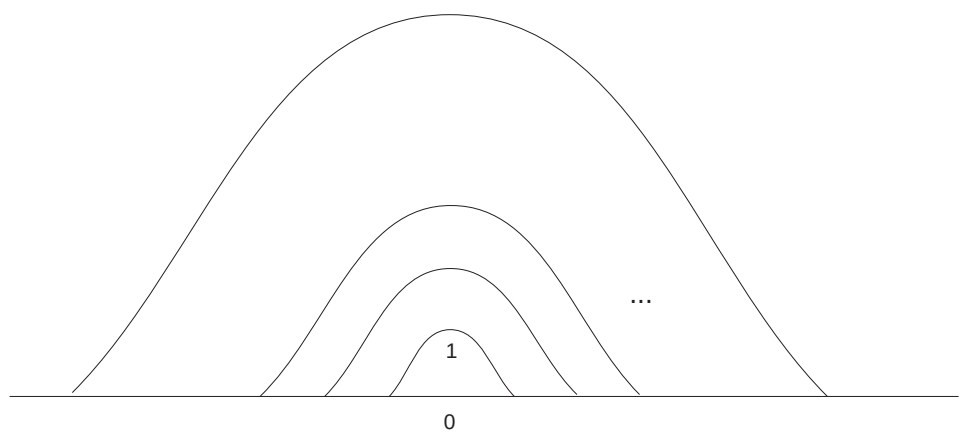

Fig.(20) The rainbow topology. Again numbers correspond to possibly different masses in the propagators.

In the spirit of the previous section we would like to define an expression of the type $[-A+B]^n(\Omega^1)$. We would expect $\Omega^1$ to be the one-loop massless two-point function, and would also need to define ${}_j\Omega^1$, its version with increased writhe number. In fact, this is indeed the correct outline for our final result, but is only so when we restrict ourselves to two-point functions which are at most linearly divergent. Consider Fig.(20) for this case and denote the graph by $\Omega^{(n)}$. Define

$$\tilde{\Omega}^{(n)} := \Omega^{(n)} \downarrow_{m_0=m_1=0} . \tag{38}$$

The notation $\ldots \downarrow_{m_0=m_1=0}$ means the following. Consider the Feynman graph in the form where all propagators have the usual quadratic denominators, and spin structures determine the numerator expression. Then set $m_0 = m_1 = 0$ in the denominator, that is nullify the masses of the inner loop. It follows that

$$< \Omega^{(n)} - \tilde{\Omega}^{(n)} >= 0, \tag{39}$$

and it is sufficient to investigate $\tilde{\Omega}^{(n)}$. We add its counterterm graphs and want then to set all masses to zero as the remaining overall divergence is mass independent. But we have to take into account that there are linear and logarithmic divergent terms to be considered (distributed over two formfactors usually) for the mass and wave-function renormalization. Therefore this problem resembles the one considered in the appendix, where the case of various form factors is studied.

The general procedure is as follows: In the numerator, keep all terms which are overall linearly or logarithmically divergent. That is, consider the numerator as a polynomial in masses and the exterior momentum. Abandon all terms which are of degree two or higher. The terms which are now linear in masses are the terms contributing to the mass renormalization, the other terms are necessarily linear in the exterior momentum at the end, so they will give the wave-function renormalization. As an example let us consider massive QED at the two-loop order.

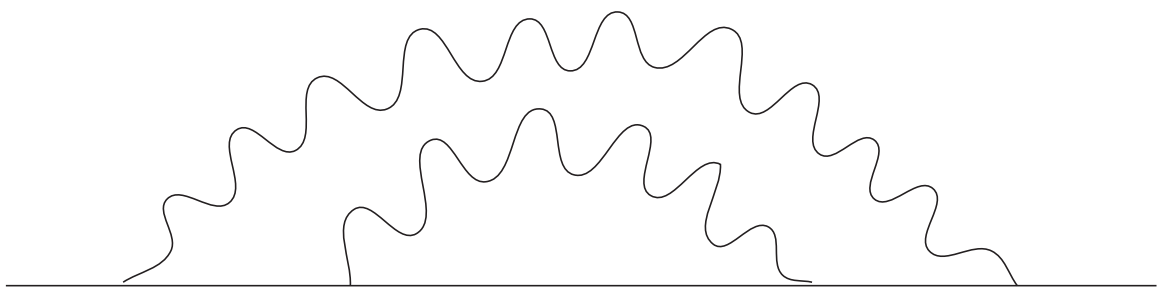

Fig.(21) An example for a mass and wave function renormalization.

The integration of the inner loop gives

$$\begin{aligned} \Omega^{(1)} &= (\Omega^1 \not{l} + m\Omega^1_m \mathbf{1})(l_2^2)^{-\varepsilon}, \\ Z_2^{(1)} &= <\Omega^1 \not{l} + m\Omega^1_m \mathbf{1}>, \\ &= Z^{(1)}_{2,w}\not{l} + Z^{(1)}_{2,m} m\mathbf{1}. \end{aligned} \tag{40}$$

Note that these $Z$-factors are not yet the standard mass and wave function renormalization.

Adding the graph and its counterterm $Z^{(1)}_{2,w}\not{l} + Z^{(1)}_{2,m} m\mathbf{1}$ gives then

$$\begin{aligned} Z_2^{(2)} : \quad & [<\Omega^1 \, {}_1\Omega^1> - << \tilde{\Omega}^1 > \Omega^1 >]\not{l}, \\ + \quad & [<2\Omega^1 \, {}_1\Omega^1_m + \Omega^1_m \, {}_1\Omega^1_m > \\ & - << 2\tilde{\Omega}^1 + \tilde{\Omega}^1 > \Omega^1 >] m\mathbf{1}. \end{aligned} \tag{41}$$

and the wave function and mass renormalization can be read off easily. The generalization to higher loop orders is obvious and follows the rule that the concatenation of terms generated by the $B$ operator might mix the different one-loop functions, as discussed in the appendix.

So far we assumed that the overall degree of divergence was not worse than linear. The case of quadratic divergences can be usually handled by appropriate differentiation with respect to the exterior momentum or by further subtractions. In the most recent method one follows the lines of [10] and subtracts in the last loop integration:

$$\frac{1}{l_n^2 - m^2} = \frac{m^4}{(l_n^2 - m^2) l_n^4} + \frac{1}{l_n^2} + \frac{m^2}{l_n^4} \tag{42}$$

The first term on the rhs has no overall degree of divergence any more and thus vanishes for the $Z$-factor contribution and the remaining terms guarantee that we can express the result in modified functions of $\Omega$-type (they might involve higher integer powers of propagators). Note that at no stage do we encounter new infrared singularities as we carefully avoid oversubtractions in this approach [10].

# 4 General Nested Topologies

So far we have considered two cases, ladder and rainbow topologies. They both give link diagrams of a simple topology, chains of pairwise concatenated links. The corresponding braid expressions are of no more than second degree in each braid group generator. Correspondingly, after applying skein relations, we have never met any knot so far. In this section we like to study the most general Feynman graphs which belong to this class. We will then show later on that exactly these Feynman graphs have a remarkable property: their contribution to the $Z$-factors is purely rational, avoiding the typical transcendental numbers appearing in loop calculations.

This allows for these transcendentals to be associated with knots, and we will show in following sections that this association is justified.

Let us combine the results of the previous two sections. We still consider vertex corrections as the basis skeleton and begin to dress internal propagators which are no rungs (so that they carry only one loop momentum in our standard assignment of loop momenta) with rainbow diagrams. It is easy to see that any other dressing would generate more complicated -knottish- link diagrams, which would bring us to braids involving higher than quadratic powers of $\sigma$'s, which we want to exclude at the moment. As we said, it will turn out that these excluded diagrams are topologically different from the simple cases considered so far. We want to understand the systematics of the simple cases first. So we are considering the following cases here:

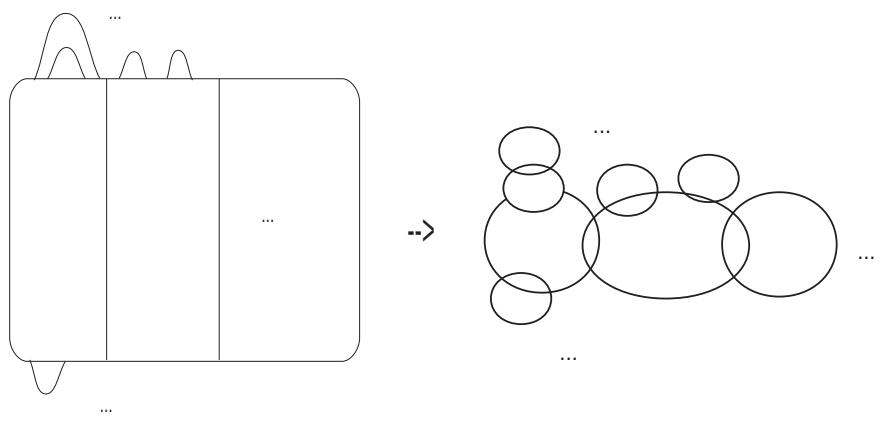

Fig.(22) Rainbow dressing of the vertex-ladder. Note that this generates only pairwise entangled link diagrams.

We want to understand how the disentanglement of disjunct subdivergences works. It is sufficient to start with the most basic example:

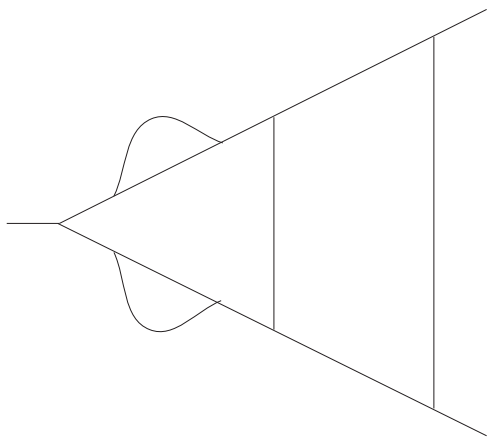

Fig.(23) A dressing generating disjunct divergences.

The dressing does not influence our argument used to reduce the skeleton to massless functions at vanishing momentum transfer. This is so because we consider not only the graph but the graph with all its counterterms. They compensate for the divergences generated by the dressing, so that the arguments used in sections two and three remain unspoiled. Thus, our basic functions $\Delta, \Omega$ encountered so far still serve as the appropriate set to express all results.

Back to our example above, renormalization theory tells us that the correct answer is

$$\begin{aligned}\text{Fig.}(25) \quad \rightarrow \quad & \Omega^1\Omega^1\ {}_2\Delta^1\ {}_3\Delta^1 - 2 < \Omega^1 > \Omega^1\ {}_1\Delta^1\ {}_2\Delta^1 \\ & + < \Omega^1 >< \Omega^1 > \Delta^1\ {}_1\Delta^1 - < \Omega^1\Omega^1\ {}_2\Delta^1 \\ & -2 < \Omega^1 > \Omega^1\ {}_1\Delta^1 + < \Omega^1 >< \Omega^1 > \Delta^1 > \Delta^1, \end{aligned} \tag{43}$$

in our notation. We see that expressions like $< \Omega^1 >< \Omega^1 >$ and $\Omega^1\Omega^1$ pop up. They are new. So far we have only met concatenated forms like $<< \Omega^1 > \Omega^1 >$ or $\Omega^1\ {}_1\Omega^1$ generated by the $A$ and $B$ operators. This was due to the total nested structure of the subdivergences considered so far. Here we have for the first time considered disjunct subdivergences. Let us compare the above example with:

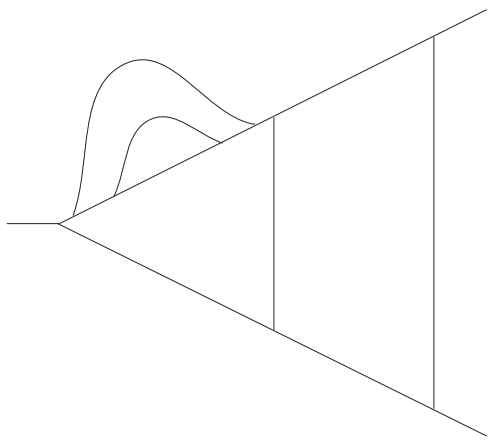

Fig.(24) Compare this nested case with the previous example.

It delivers

$$\begin{aligned}\text{Fig.}(24) \quad \rightarrow \quad & \Omega^1\ {}_1\Omega^1\ {}_2\Delta^1\ {}_3\Delta^1 - < \Omega^1 > \Omega^1\ {}_1\Delta^1\ {}_2\Delta^1 \\ & - < \Omega^1\ {}_1\Omega^1 > \Delta^1\ {}_1\Delta^1 + < \Omega^1 < \Omega^1 >> \Delta^1\ {}_1\Delta^1 \\ & - < \Omega^1\ {}_1\Omega^1\ {}_2\Delta^1 - < \Omega^1 > \Omega^1\ {}_1\Delta^1 \\ & - < \Omega^1\ {}_1\Omega^1 > \Delta^1 + < \Omega^1 >< \Omega^1 > \Delta^1 > \Delta^1, \end{aligned} \tag{44}$$

which is familiar from sections two and three.

The reason for the difference is obvious. In the first case we have disjunct forests at the skeleton (the two one-loop self-energies) while in the second case we have the completely nested two-loop self-energy.

We have to transfer this information to the link diagrams and must find a notation for it. What we want to do is dress the link diagram corresponding to the dressing of the skeleton graph. To this end let us label the propagators once and for all:

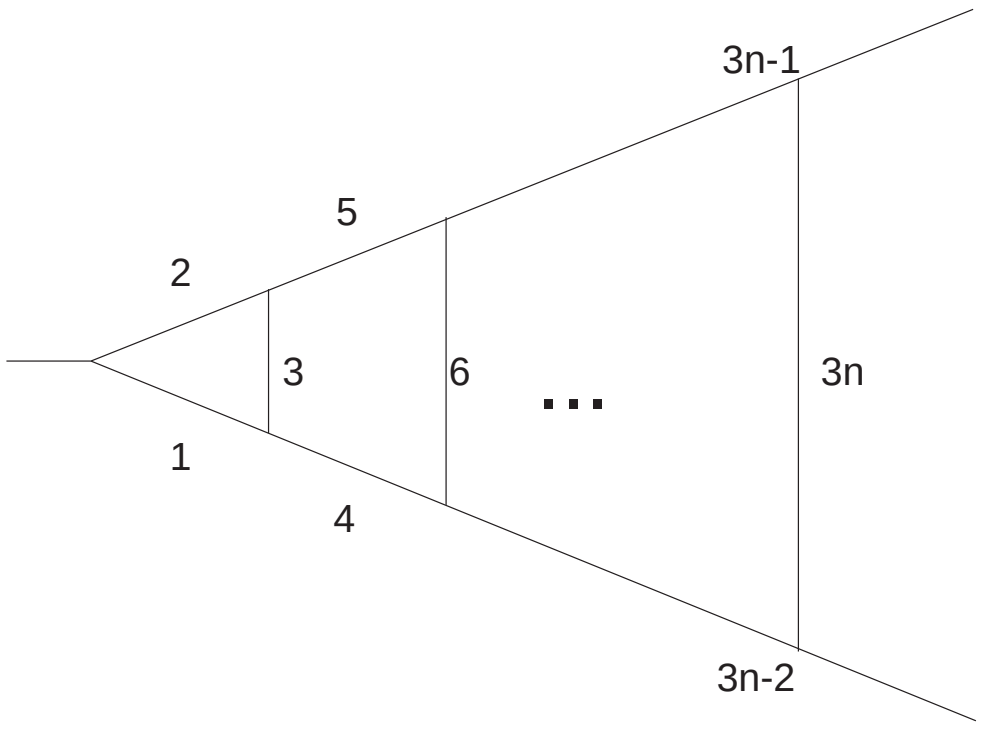

Fig.(25) House-numbering of our skeleton propagators. The propagators $1, 2, 4, 5, 7, 8, \ldots$ can be dressed with rainbow topologies.

So, we are considering rainbow dressing of the propagators $(1, 4, \ldots)$ or $(2, 5, \ldots)$ in the vertex case. In principle we can describe then every dressed propagator by giving its housenumber and the loop number of the rainbow diagram dressing it, and can iteratively continue dressing propagators in rainbows and so on, still remaining in link diagrams only having pairwise connected components. They all correspond as braids to products of the very simple topology $\prod_{i=1}^{r} \sigma_i^2$.

Now following the rule that the $A$ and $B$ actions in the skein relation concatenate the satellite links towards the skeleton link we get correct results:

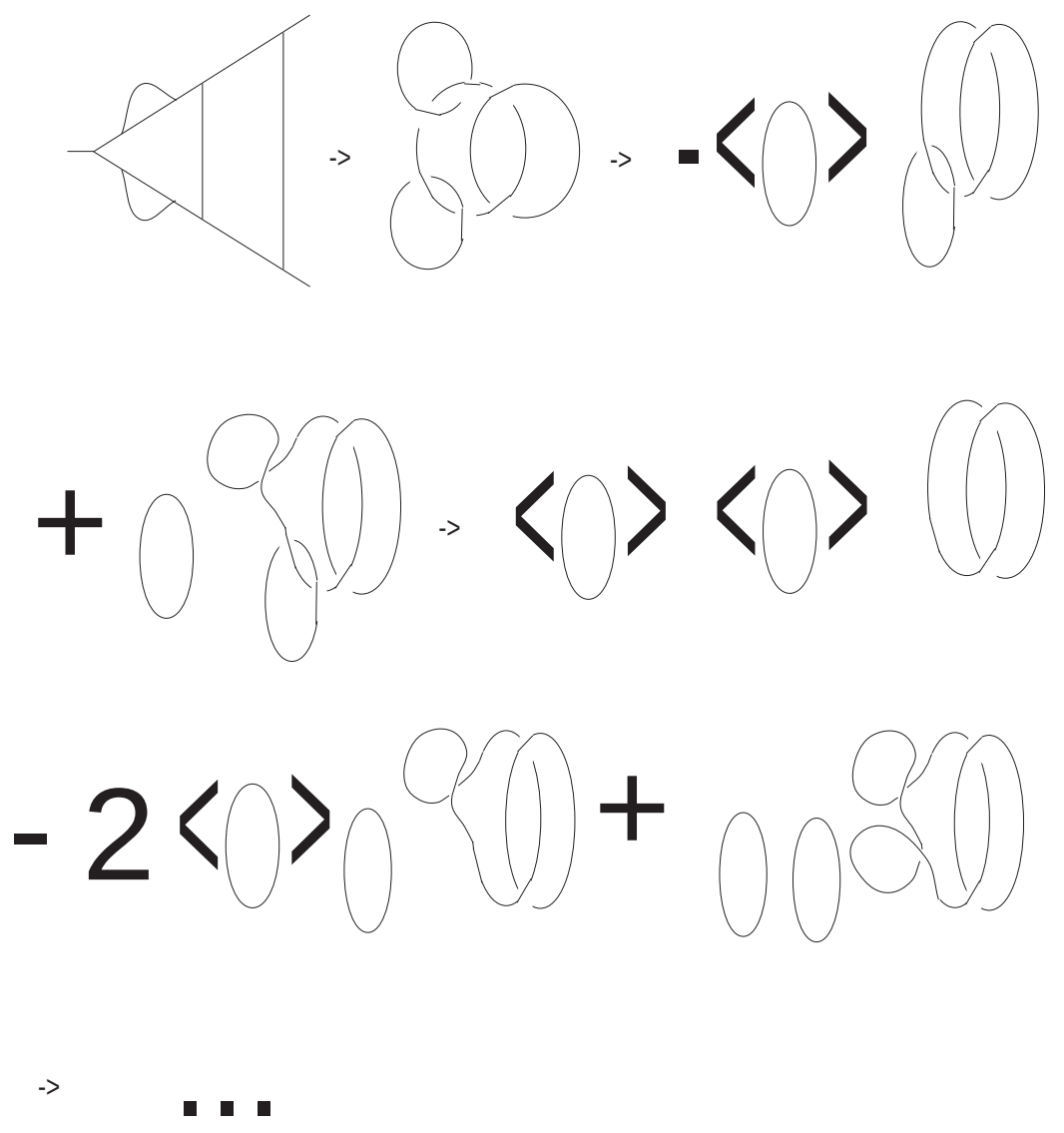

Fig.(26) An example calculation using our graphical notation to demonstrate the behaviour of $A$ and $B$ on disjunct subdivergences.

We see the underlying principle: $A$ and $B$ treat disjunct subgraphs as disjunct, that is they factorize instead of concatenate the corresponding one-loop functions: $\Omega^1\Omega^1$ instead of $\Omega^1{}_1\Omega^1$ and $<\Omega^1><\Omega^1>$ instead of $<<\Omega^1>\Omega^1>$.

We finish our considerations of nested and disjunct divergences and turn to overlapping divergences. We hope to find a similar structure for topological simple graphs there. This is crucial for our final attempt to identify the topological nature of a Feynman graph with certain properties of its divergent part.

# 5 Overlapping Divergences

In this section we will ourselves address to overlapping divergences. We will study graphs of the form:

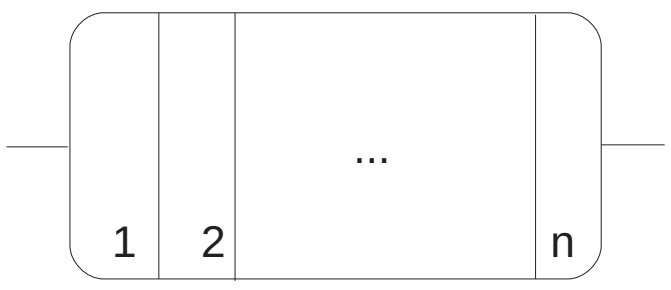

Fig.(27) The overlapping ladder. These link topologies are still simple.

and their cable generalizations:

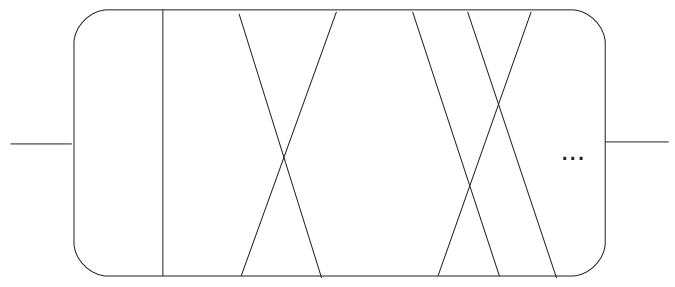

Fig.(28) The generalized case including various topologies for the rungs.

We consider these type of graphs together with the ones of section 2 and 3 as the skeletons for later purposes. It is the main objective of this section to show that also the overlapping divergences factorize in a manner similar to the cases studied so far. This then allows to classify them by their corresponding link algebra too. The main result of this section is that the overlapping ladder topology gives the same concatenations as before, with the only modification that we have to sum over all possibilities to identify subdivergences in the graph. As a consequence, this allows us in later subsections to establish rationality also for simple topologies in the case of overlapping divergences. This in turn defines transcendentality of $Z$-factors as a sensible test for knots, that is non-trivial topologies.

Let us start with a simple two-loop example. The graph has two overlapping subdivergences, so we have to calculate the following expressions

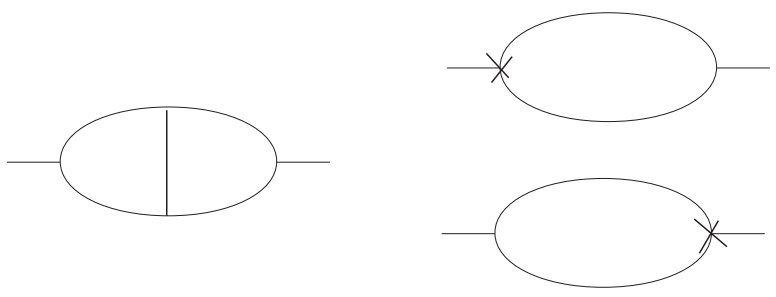

Fig.(29) The two-loop example.

In the denominator we have five propagators. Let $N_i$ the product of all propagators involving the loop momentum $l_i$. It is implicitly understood that there is some numerator taking particle type and spin structures into account, but we operate in the following only on the scalar denominator part of the Green function under consideration. Nevertheless we remind the reader that the numerator structure affects the power counting, so that in the following one cannot use the exlicitly given denominator expressions for power counting purposes. In the above example we have $N_l = P_1P_2P_3$, $N_k = P_3P_4P_5$. We define

$$\begin{aligned} \tilde{N}_i &:= N_i|_{q=0,m_j=0\ \forall j}, \\ \mathbf{N_i} &:= \frac{\tilde{N}_i - N_i}{\tilde{N}_i}. \end{aligned} \tag{45}$$

Then we have

$$\frac{1}{P_1P_2P_3P_4P_5} = \frac{1}{P_1P_2P_3P_4P_5}\mathbf{N_l}\mathbf{N_k} - \frac{1}{\tilde{N}_lP_4P_5} - \frac{1}{P_1P_2\tilde{N}_k} + \frac{P_3}{\tilde{N}_l\tilde{N}_k},$$

$$\text{implying}\quad \Omega^{(2)} \equiv \Omega^{(2)}_f - \Omega^{(2)}|_l - \Omega^{(2)}|_k + \Omega^{(2)}|_{l,k}, \tag{46}$$

by construction. The last term on the rhs vanishes in DR. This poses no serious problem as we only have to use the relation

$$\int d^D l (l^2)^{-r} = 0, \tag{47}$$

for cases $(r - D) \neq \varepsilon$.

Let us now investigate Eq.(46). The first term on the rhs is UV-convergent, as long as our overall degree of divergence was not worse than linear; this is always the case. We have to be cautious about the quadratic divergences of boson propagators. But they are usually safe by gauge invariance, which reduces the overall degree of divergence to a logarithmic degree. Otherwise, we would follow the recipe of section 2 and improve the powercounting by further subtractions.

Back to our considerations of Eq.(46) we notice that the first term on the rhs will not contribute to our MS $Z$-factor, due to its UV convergence:

$$< \frac{1}{P_1 P_2 P_3 P_4 P_5} \mathbf{N_l N_k} >= 0. \tag{48}$$

The remaining terms to be considered are the second and third term on the rhs of Eq.(46). The counterterms contribute

$$-Z_1^{(1)} \int d^D l \frac{1}{P_4 P_5} - Z_1^{(1)} \int d^D l \frac{1}{P_1 P_2}; \tag{49}$$

adding these contributions to the remaining two terms on the rhs of Eq.(46) gives us a result free of subdivergences. So we can set all remaining masses to zero in this sum. We then have $P_1 = P_5, P_2 = P_4$. Our UV-divergences are thus contained in

$$\begin{aligned} Z_2^{(2)} &= 2 < \Delta^1 \int d^D l \frac{(l^2)^{-\varepsilon}}{P_1|_{m_1=0} P_2|_{m_2=0}} \\ &- < \Delta^1 > \int d^D l \frac{1}{P_1|_{m_1=0} P_2|_{m_2=0}} > . \end{aligned} \tag{50}$$

Defining

$$\begin{aligned} {}_j\Omega^1 &:= \int d^D l \frac{(l^2)^{-\varepsilon j}}{P_1|_{m_1=0} P_2|_{m_2=0}}, \\ \Omega^1 &:= {}_0\Omega^1, \end{aligned} \tag{51}$$

we have

$$Z_2^{(2)} = 2 < \Delta^1 \, {}_1\Omega^1 - < \Delta^1 > \Omega^1 >, \tag{52}$$

which has a striking similarity with our result Eq.(17). The main difference is the factor of two which reflects the overlapping structure in this simple example.

Here, $Z_2^{(2)}$ stands for the two-loop divergences of Fig.(29). It is a condensed notation for the two $Z$-factors corresponding to a two-point function usually, the mass, $Z_{2,m}$, and wave-function renormalization, $Z_{2,w}$. One would extract them by looking for their corresponding form factors, e.g. $\not{q}$ and $m\mathbf{1}$ for the fermion propagator, and by Taylor expanding at the mass-shell. Graphically, our result is:

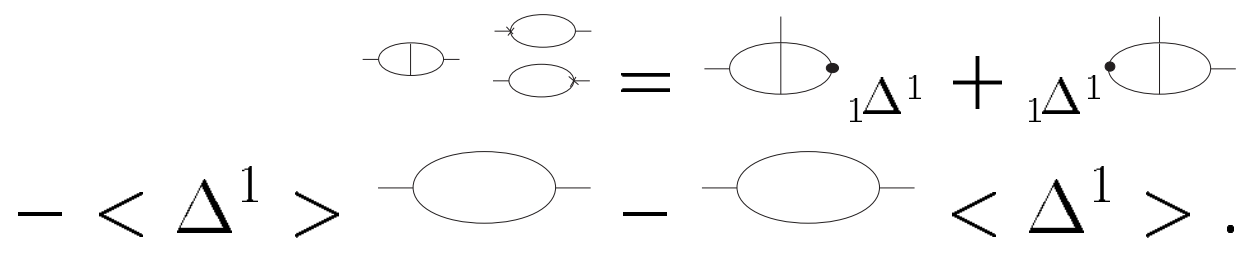

Fig.(30) Some notation emphasizing the fact that the result was just a sum over all possibilities to interprete one loop as the skeleton ($\Omega$) and the other one as the subdivergence ($\Delta$).

We want to generalize this approach to the $n$ loop case. We are looking for a general prescription to convert overlapping topologies back to products of one-loop graphs, as we did before for the nested topologies. We then hope to find a similar correspondence to braid structures. Consider the $n$-loop ladder graph Fig.(27). We claim that

$$\begin{aligned}\Omega^n_f &= \Omega^n - \Omega^n|_{l_1}\Omega^n|_{l_n} + \Omega^n|_{l_1,l_n}\\ &= \Omega^n N_{l_1} N_{l_n},\end{aligned}\qquad(53)$$

is finite. The functions $\Omega^n|_{\ldots}$ used above are implicitly defined by the second line in the above equation. Note that $\Omega^n|_{l_1,l_n}$, the last term on the rhs in the first line, does not vanish if $n \neq 2$. The assertion on $\Omega^n_f$ is justified by the fact that all possible divergent sectors cancel in the above expression. This can also easily seen by do a powercounting for Eq.(53) or by investigating the Dyson Schwinger equations.

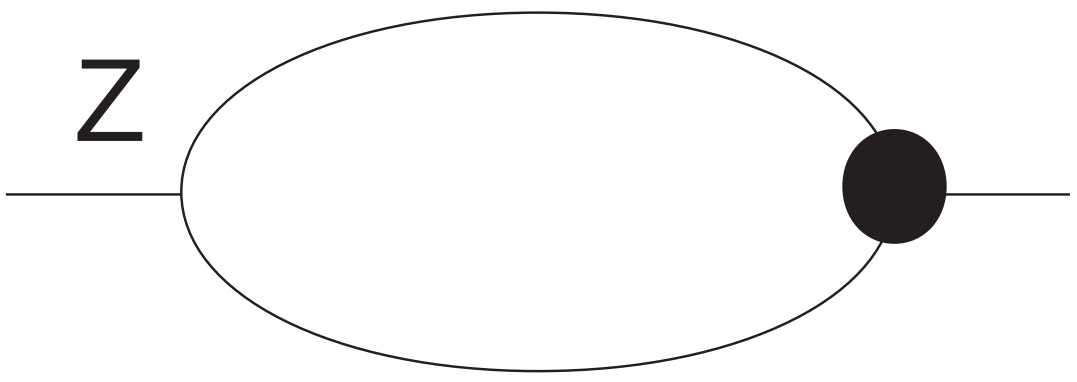

Fig.(31) The Dyson Schwinger equations for a propagator. Our subtractions in Eq.(50) on the lhs ($l_1$) and rhs ($l_n$) remove the divergences. So the $Z$ factor in the DS equation becomes trivial ($Z = 1$), and the blob has a subtracted form, cf. section 2.

This is in agreement with the general result that an overlapping divergence renormalizes by subtracting out all divergent subgraphs in a manner different from disjunct or nested divergences [13]. While for these cases also the product of the operators removing the subdivergences will appear, in the overlapping case this product structure is not maintained. It is replaced by a sum over subtractions at all divergent subgraphs.

Therefore, what remains to be calculated are the last three terms on the rhs of Eq.(53).

$$< \Omega_f^n >= 0 \Rightarrow < \Omega^n >=< \Omega^n|_1 + \Omega^n|_n - \Omega^n|_{1,n} > \tag{54}$$

Let us add our counterterm expressions.

$$\begin{aligned} &-2 \sum_{i=1\ldots n-1} Z_1^{(i)} \Omega^{(n-i)} \\ +2 &\sum_{\substack{i=1\ldots n-1 \\ j=1\ldots n-1 \\ 1<i+j<n,\, i\neq j}} Z_1^{(i)} Z_1^{(j)} \Omega^{(n-i-j)} \\ &+ \sum_{i=1\ldots n-1} (Z_1^{(i)})^2 \Omega^{(n-2i)} \end{aligned} \tag{55}$$

The whole contribution then is, in our graphical notation:

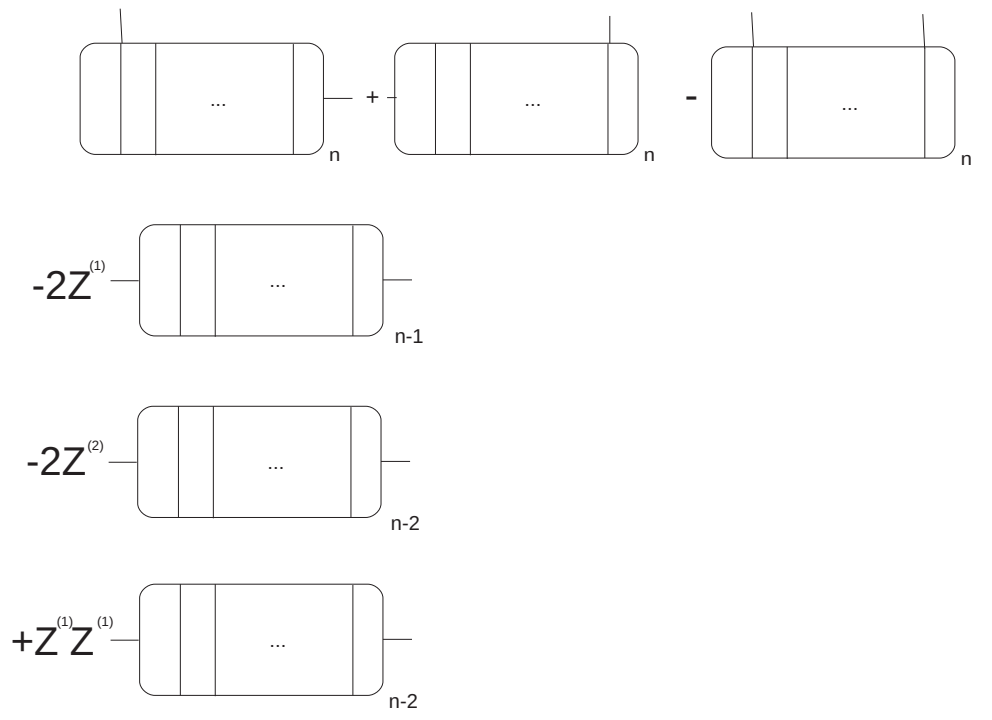

Fig.(32) Our ladder with all its counterterms. In the first line wee see the result of Eq.(54).

This expression is free of subdivergences and we can set all remaining masses to zero. This is still not quite what we want as expressions like still involve loops which are not expressible in terms of our basic functions $\Delta^1, \Omega^1$. Only expressions where the exterior momentum flows only through one loop are amenable to our procedure.

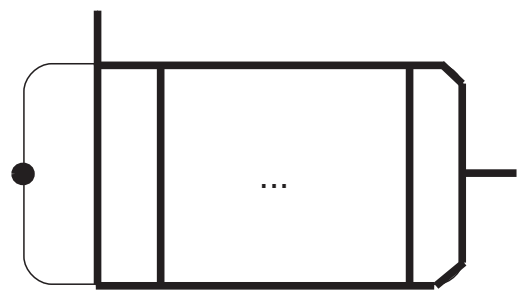

Fig.(33) These expressions still involve functions which do not factorize into our basic $\Delta^1$'s and $\Omega^1$'s. Typically, the expressions involve three-point functions to be evaluated at non-vanishing momentum transfer, drawn in thick lines above.

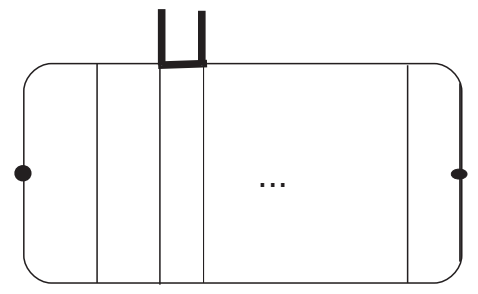

Fig.(34) These '1-state' expressions are easy to calculate. The thick line indicates the only propagator carrying the external momentum $q$.

On the other hand, we know that in the above sum including all counterterms only local contributions can survive, demanding that there are cross cancellations between the various terms.

To see how we can proceed let us recapitulate what we have done in Eq.(53). We wrote the original function $\Omega^{(n)}$ as a difference between a function which was UV-convergent $\Omega_f^{(n)}$ and some simpler functions $\Omega^{(n)}|_i$. The guiding principle was Weinberg's theorem [6]. In the form which is useful for us it states that once a graph is finite by powercounting for its overall degree of divergence and all its subdivergences we can be sure that it is convergent. This allowed us to conclude that $<\Omega_f^{(n)}>= 0$. But Weinberg's theorem also tells us that an expression is convergent if it has a vanishing overall degree of divergence and either it has no subdivergences (the previous case) or it has all its subdivergences subtracted by appropriate counterterms. This statement is just the underlying principle which determines the counterterms required to make a renormalizable theory finite. Vice versa, we know that for every combination of loop momenta providing a subdivergence, there exist an appropriate counterterm in the sum Eq.(55). We will show that this gives us an algorithm to continue the process which was started in Eq.(53) to simplify $\Omega^{(n)}$. But we will give the result first. To this end let us introduce some notation.

- The massless $n$-loop ladder graph: $^{|}\Box \ldots \Box^{|}_n$, where $^{|}$ indicates the flow of the external momenta. We call this a $(0,n)_n$ state, as the momenta flows through the whole $n$-loop graph.

- The $(i,j)_n$ state: $\Box\ldots\Box^{|}_{i}\Box\ldots\Box^{|}_{j}\Box\ldots\Box_n$, we agree to count $i$ from the left and $j$ from the right.
- Counterterm graphs: $\Box\ldots\Box_i]\Box\ldots\Box^{|}_{r}\Box\ldots\Box^{|}_{s}\Box\ldots\Box_{n-i-j}[\Box\ldots\Box_j$ for a $(n-i-j)$-loop graph in a $(r,s)_{n-i-j}$ state multiplied by $Z^{(i)}Z^{(j)}$. Note that this is a condensed notation for a whole $Z$-factor, e.g.: $\Box\Box] = (\Box\Box - \Box]\Box)]$.

Our final claim is that the following expression, involving 1-states only, gives the correct result for the $Z_2$-factor.

$$
\begin{aligned}
& \sum_{i=0}^{n} \Box\ldots\Box^{|}_{i}\Box^{|}\Box\ldots\Box_n \\
-\Box] & \sum_{i=0}^{n} \Box\ldots\Box^{|}_{i}\Box^{|}\Box\ldots\Box_{n-1} \\
- & \sum_{i=0}^{n} \Box\ldots\Box^{|}_{i}\Box^{|}\Box\ldots\Box_{n-1}\,[\Box \\
-\Box\Box] & \sum_{i=0}^{n} \Box\ldots\Box^{|}_{i}\Box^{|}\Box\ldots\Box_{n-2} \\
- & \sum_{i=0}^{n} \Box\ldots\Box^{|}_{i}\Box^{|}\Box\ldots\Box_{n-2}\,[\Box\Box \\
& \ldots \\
+\Box] & \sum_{i=0}^{n} \Box\ldots\Box^{|}_{i}\Box^{|}\Box\ldots\Box_{n-2}\,[\Box \\
& \ldots
\end{aligned}
\qquad (56)
$$

Now, to prove it as advocated above, we use the fact that an overall convergent expression, when dressed with internal vertex and self-energy corrections, can be rendered finite by including the appropriate counterterms. So what we have to do is to do the step in Eq.(53) at the same time for a $n$-loop state and some appropriate chosen counterterm states so that in each step all subdivergences are compensated.

We then want to repeat the step in Eq.(53) $n-1$ times so that we end up with expressions where the exterior momenta flows only through one loop momentum -1-states, that is:

$$
\begin{aligned}
& \Box\ldots\Box^{|}_{i}\Box^{|}\Box\ldots\Box_n \\
= {} & B^{i-1}(\Delta^1)B^{n-i-2}(\Delta^1)\;{}_{n-1}\Omega^1.
\end{aligned}
$$

These 1-states are easy to calculate. All loops which are free of the exterior momentum correspond to functions ${}_j\Delta^1$ and the exterior momentum flow is in ${}_{n-1}\Omega^1$. So we obtain the usual concatenations with the $B$ operator. In the spirit of Eq.(53) we would like to write

$$\Omega^{(n)}|_1 \quad = \quad \Omega^{(n)}|_1 N_{l_2} N_{l_n} + \Omega^{(n)}|_{1,2} + \Omega^{(n)}|_{1,n} - \Omega^{(n)}|_{1,2,n}. \tag{57}$$

Here $|_1$ denotes the loop which is free of the external momentum already. If we now could push the analogy to Eq.(53) further and conclude

$$< \Omega^{(n)}|_1 N_{l_2} N_{l_n} = 0 > ? \tag{58}$$

we would see the beginning of an algorithm leading to the final state in Eq.(56). The problem is that now, in Eq.(57), we have removed the overall degree of divergence but not all subdivergences. We have removed all subdivergences not involving $l_1$, but the loop corresponding to $l_1$ produces a problem we have not taken care of yet. We can do so by remembering Weinberg's theorem and picking up the appropriate counterterm expression:

$$< \Omega^{(n)}|_1 N_{l_2} N_{l_n} - Z_1^{(1)} \Omega^{(n-1)}|_1 N_{l_2} N_{l_n} >= 0, \tag{59}$$

where $\Omega^{(n-1)}$ was expressed in $n-1$ loop momenta $l_2, \ldots, l_n$. We have in our $\Box$ notation:

$$\begin{aligned}
&< \Box^{|}_{1}\Box \ldots \Box^{|}_{n} \;-\; \Box]\,{}^{|}\Box \ldots \Box^{|}_{n-1} > \\
=&< \Box\Box^{|}_{2}\Box \ldots \Box^{|}_{n} \;-\; \Box]\,\Box^{|}_{1}\Box \ldots \Box^{|}_{n-1} > \\
+&\;\; < \Box^{|}_{1}\Box \ldots \Box^{|}_{n-1}\Box_n \;-\; \Box]\,{}^{|}\Box \ldots \Box^{|}_{n-2}\Box_{n-1} > \\
-&\;\; < \Box\Box^{|}_{2}\Box \ldots \Box^{|}_{n-1}\Box_n \;+\; \Box]\,\Box^{|}_{1}\Box \ldots \Box^{|}_{n-2}\Box_{n-1} > .
\end{aligned}$$

Applying the above mechanism $n-1$ times for a $n$-loop graph gives us the sum over 1-states predicted in the final result above. At each step, by the very definition of renormalizability, there is an appropriate set of counterterms available so that the mechanism is justified. It is easy to see that the signs in the relation Eq.(57) conspire in the right way to guarantee that each 1-state appears exactly one time in the final sum.

We give an example for the case $n = 3$:

$$\begin{aligned}
& {}^{|}\Box\Box\Box^{|} \\
& - \Box]\,{}^{|}\Box\Box^{|} - {}^{|}\Box\Box^{|}[\Box \\
& - \Box\Box]^{|}\Box^{|} - {}^{|}\Box^{|}[\Box\Box + \Box]^{|}\Box^{|}[\Box \\
& = {}^{|}\Box\Box^{|}\Box + \Box^{|}\Box\Box^{|} - \Box^{|}\Box^{|}\Box \\
& - \Box]^{|}\Box\Box^{|} - {}^{|}\Box\Box^{|}[\Box \\
& - \Box\Box]^{|}\Box^{|} - {}^{|}\Box^{|}[\Box\Box + \Box]^{|}\Box^{|}[\Box \\
& = {}^{|}\Box^{|}\Box\Box + \Box^{|}\Box^{|}\Box - \Box^{||}\Box\Box \\
& - \Box^{|}\Box^{|}[\Box - {}^{|}\Box^{|}\Box[\Box + \Box^{||}\Box[\Box \\
& + \Box\Box^{|}\Box^{|} + \Box^{|}\Box^{|}\Box - \Box\Box^{||}\Box
\end{aligned}$$

$$
\begin{aligned}
&- \square]^{|}\square^{|}\square \;-\; \square]\square^{|}\square^{|} \;+\; \square^{||}\square[\square \\
&- \square\square]^{|}\square^{|} \;-\; {}^{|}\square^{|}[\square\square \;+\; \square]^{|}\square^{|}[\square \\
&= {}^{|}\square^{|}\square\square \;+\; \square^{|}\square^{|}\square \;+\; \square\square^{|}\square^{|} \\
&- \square]({}^{|}\square^{|}\square \;+\; \square^{|}\square^{|}) - ({}^{|}\square^{|}\square \;+\; \square^{|}\square^{|})[\square \\
&- \square\square]^{|}\square^{|} \;-\; {}^{|}\square^{|}[\square\square \\
&+ \square]^{|}\square^{|}[\square.
\end{aligned}
$$

Here we used $\square^{||}\square = 0$, as these expressions correspond to graphs where the exterior momentum does not flow at all through the graph. For a massless graph in DR we have then a vanishing tadpole graph, cf. Eq.(47).

Now let us derive the same result from knot theory. Let us start again with the two-loop example. According to our rules in the previous section we obtain:

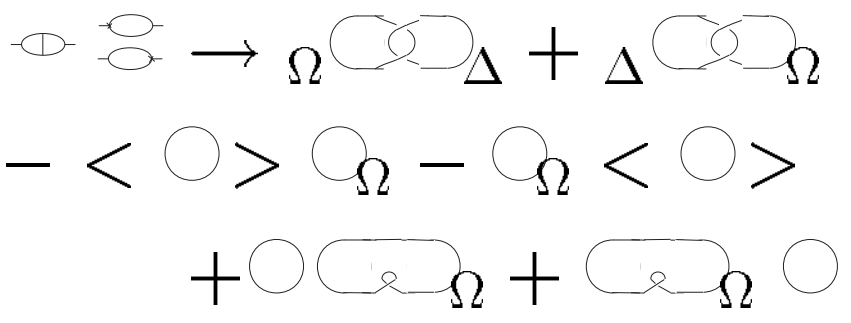

This time the correspondence between a Feynman graph and a link diagram involves a sum over possibilities opened up by the overlapping topology. We have not changed the rules concerning overcrossings, but now one link will correspond to a vertex correction $\Delta^1$, the other link to a self-energy $\Omega^1$. The first question we have to answer is: which link serves as the skeleton and which one as the subdivergence? In case of the nested topologies considered so far we could always decide this in a unique manner. By the very definition of a overlapping divergence both links above can play the role of the skeleton or subdivergence.

Accordingly, as indicated above, let us sum over both possibilities. Applying now the formalism developed in the previous section, taking into account that the skeleton is necessarily a two point function, the subdivergence necessarily a vertex correction, we obtain

$$
\begin{aligned}
Z_2^{(2)} &= -<\Delta^1>\Omega^1+\Delta^1\,{}_1\Omega^1 - <\Delta^1>\Omega^1+\Delta^1\,{}_1\Omega^1 \\
&= 2(-<\Delta^1>\Omega^1+\Delta^1\,{}_1\Omega^1), \qquad (60)
\end{aligned}
$$

which is the desired answer. Encouraged by this result we investigate the three-loop case:

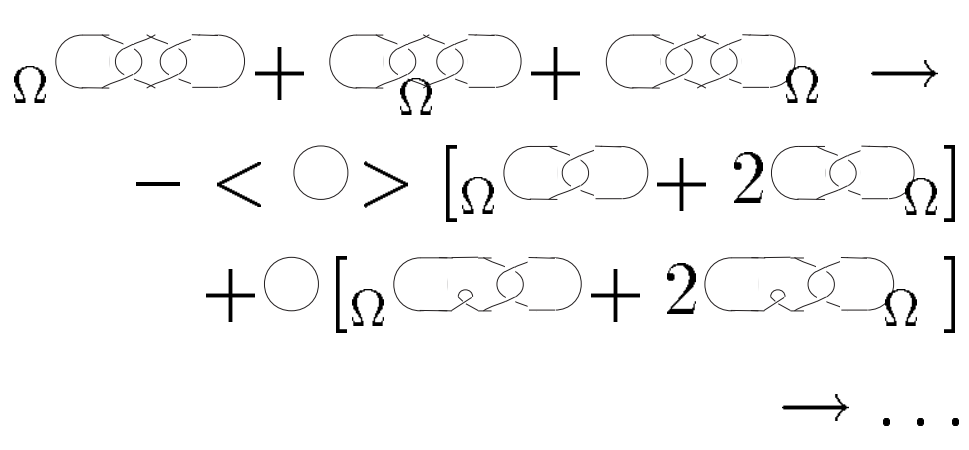

Fig.(35) A three-loop example.

We indicated the skeleton loop by an $\Omega$. Explicitly in our one-loop functions we have

$$\begin{aligned}
Z_2^{(3)} &= 2[-<\Delta^1>\Delta^1\;{}_1\Omega^1 - <\Delta^1\;{}_1\Delta^1>\Omega^1 \\
&\quad + <<\Delta^1>\Delta^1>\Omega^1 + \Delta^1\;{}_1\Delta^1\;{}_2\Omega^1] \\
&\quad -2<\Delta^1>\Delta^1\;{}_1\Omega^1 + <\Delta^1><\Delta^1>\Omega^1 + \Delta^1\Delta^1\;{}_2\Omega^1 \\
&= [2\Delta^1\;{}_1\Delta^1\;{}_2\Omega^1 + \Delta^1\Delta^1\;{}_2\Omega^1] \\
&\quad -[4<\Delta^1>\Delta^1\;{}_1\Omega^1] \\
&\quad -2[<\Delta^1\;{}_1\Delta^1>\Omega^1 - <<\Delta^1>\Delta^1>\Omega^1] \\
&\quad + <\Delta^1><\Delta^1>\Omega^1. \qquad (61)
\end{aligned}$$

Note that the case where we read the three-loop overlapping ladder as having a one-loop correction on each side, is the case of two disjunct subdivergences. That was the reason why we discussed disjunct subdivergences in some detail in the previous section.

A comparison with Eq.(56) shows that it is again the correct result. Indeed, the sum over all possible assignments of the skeleton property to one ring equals the sum over all possibilities which ring will carry the external momentum flow, and we see that the knot theoretic approach gives the correct answer immediately.

In analogy to the previous sections we generalize this result to the case of various topologies. We want to be able to calculate cases like

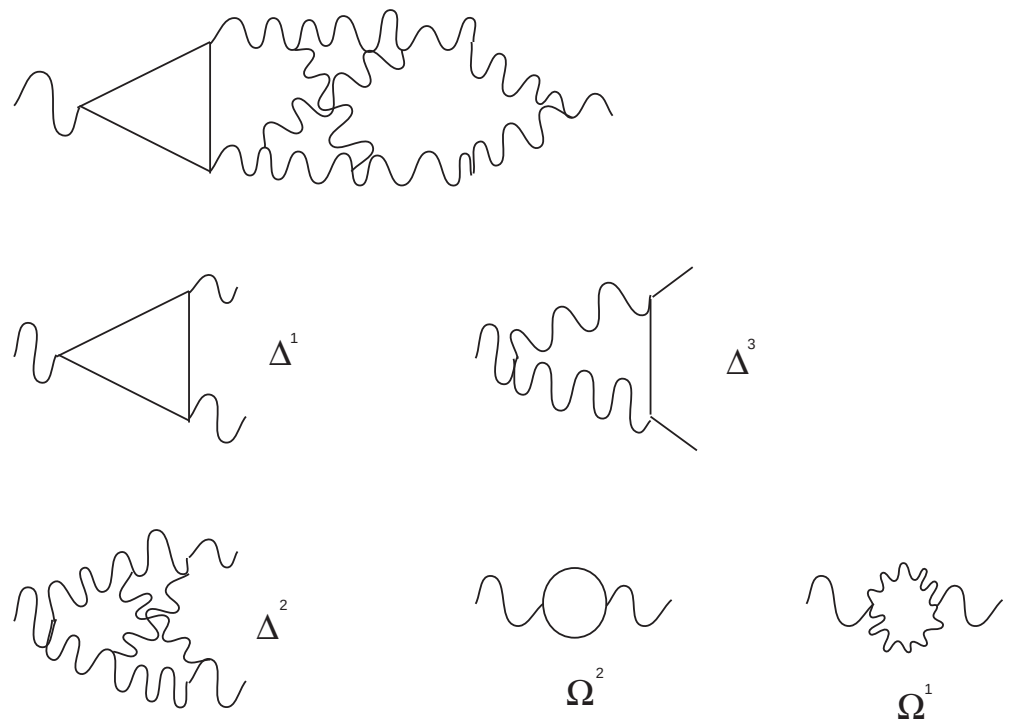

Fig.(36) A more general case which demands a proper notation vor various types of diagrams and topologies.

Restricting ourselves to the transversal part $\sim g_{\mu\nu}q^2 - q_\mu q_\nu$ (to avoid further subtractions in the $\Omega$'s, for the longitudinal part we would have to use a derivative with respect to the external momentum, see section 3) which is only overall logarithmically divergent, we would find as the result in an obvious generalization of the results of Eq.(56):

$$\begin{aligned} Z_2(\mathrm{Fig.}(52)) \;=\; & [\Delta^1 \,{}_1\Delta^2 \,{}_3\Omega^1 + \Delta^2 \,{}_2\Delta^3 \,{}_3\Omega^2 \\ & +\Delta^1\Delta^2 \,{}_3\Omega^1] \\ & -[<\Delta^1>\Delta^2 \,{}_2\Omega^1 - <\Delta^2>\Delta^3 \,{}_1\Omega^2 \\ & - <\Delta^1>\Delta^2 \,{}_2\Omega^1 - <\Delta^2>\Delta^1 \,{}_1\Omega^1] \\ & -[<\Delta^1 \,{}_1\Delta^2>\Omega^1 - <<\Delta^1>\Delta^2>\Omega^1 \\ & + <\Delta^2 \,{}_2\Delta^3>\Omega^2 - <<\Delta^2>\Delta^3>\Omega^2] \\ & + <\Delta^1><\Delta^2>\Omega^1. \end{aligned} \tag{62}$$

These are the 12 expected terms ($2^{n-1} = 4$, from the skein relation, times 3 possibilities to assign the skeleton property).

In case that the incoming particle and the outgoing particle are different this would blow up the possibilities, e.g:

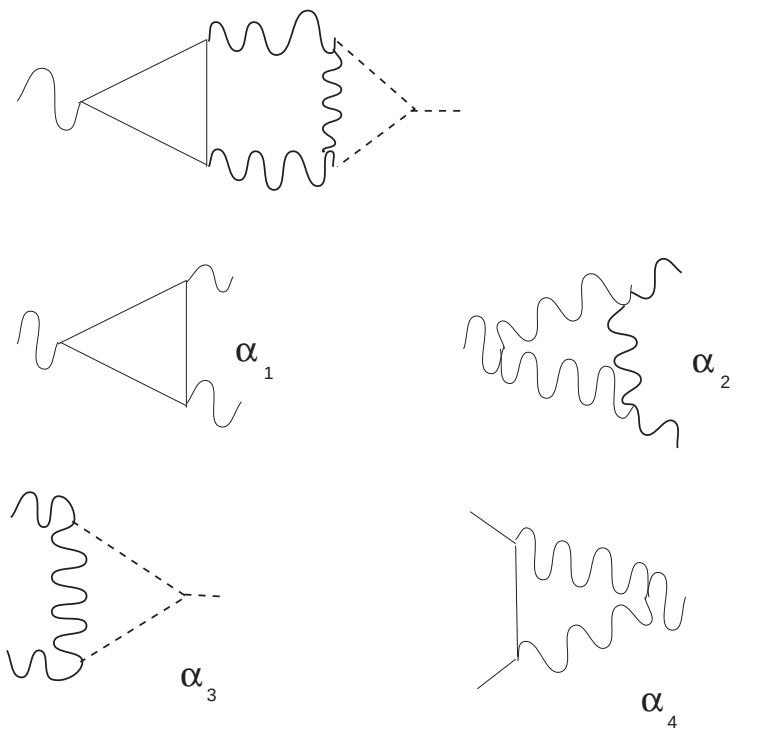

Fig.(37) From the left and from the right we get different $\alpha$'s.

This completes our treatment of overlapping topologies. We succeeded in representing a sufficient large class of Feynman graphs in terms of concatenated one-loop functions. Let us see now what we can learn from these structures.

# 6 Rational Divergences

We are now in a position to investigate all Feynman graphs which correspond to simple topologies. We define a topology as simple when it corresponds to a braid group expression of at most second degree in all braid generators. These braids, after applying the skein relation $n$-times, will generate an unknot with writhe number $n$. This property of being free of knots should reflect itself in the $Z$-factors, when we want to have a chance to identify $Z$-factors with topology -knots- in some way. We claim that there is a one-to-one correspondence between the transcendental numbers in the divergent part of a diagram and the knot in its link diagram, so we better prove that such transcendentals do not pop up in simple topologies.

To give the reader an idea what sort of cancellations are necessary we give in the appendix a result for a seven loop ladder topology. The graph itself has transcendentals in abundance in its divergent part, but when we add its counterterm expressions they all disappear.

We proceed in the following way. We first investigate how this cancellation of transcendentals appears for some special choice (ladder topology in a scalar theory) of the $\Delta$ function. In fact, we show that for an arbitrary loop order $n$ the highest possible transcendental $\zeta(n-1)$ will not pop up. The proof for the other transcendentals is similar.

We indicate how the proof has to be generalized for the arbitrary tensor case.

Then we discuss this cancellation of transcendentals from a more general viewpoint and give a general inductive proof.

The result of this section is the following statement: Any Feynman graph in a renormalizable theory which corresponds to a simple topology in its link diagram expression gives rise to only rational divergences when calculated together with its counterterm graphs. Here rational divergences means rational numbers as coefficients of the proper Laurent part in the DR expansion parameter $\varepsilon$.

Let us start considering the following function, which exhibits all typical properties of the observed cancellations:

$$\begin{aligned} {}_j\Delta : \quad & \int d^Dk \frac{(k^2)^{-\varepsilon j}}{k^2(k+q)^2} =: (q^2)^{-\varepsilon(j+1)} {}_j\Delta, \\ \Rightarrow {}_j\Delta \quad = \quad & \frac{\Gamma(1+(j+1)\varepsilon)\Gamma(1-\varepsilon)\Gamma(1-(j+1)\varepsilon)}{(j+1)\varepsilon(1-(j+2)\varepsilon)\Gamma(1+j\varepsilon)\Gamma(1-(j+2)\varepsilon)}. \end{aligned}$$

Define

$$\begin{aligned} P_n \quad & := \quad \prod_{i=0}^{n-1} {}_i\Delta, \\ \Rightarrow P_n \quad & = \quad \frac{\Gamma(1-\varepsilon)^{n+1}\Gamma(1+n\varepsilon)}{n!\varepsilon^n(1-2\varepsilon)\ldots(1-(n+1)\varepsilon)\Gamma(1-(n+1)\varepsilon)}. \end{aligned}$$

Now use

$$\Gamma(1-z) = \exp(\gamma z)\exp(\sum_{j=2}^{\infty} \frac{\zeta(j)}{j} z^j). \tag{63}$$

It follows

$$\begin{aligned} P_n \quad = \quad & \frac{1}{n!\varepsilon^n(1-2\varepsilon)\ldots(1-(n+1)\varepsilon)} \exp(-n\gamma\varepsilon) \\ & \exp(\sum_{j=2}^{\infty} \frac{\zeta(j)}{j} \varepsilon^j [n+1+(-n)^j-(n+1)^j]). \end{aligned}$$

We conclude immediately that $\zeta(2)$ can not appear in a $Z$-factor contribution as its coefficient is $(n+1+n^2-(n+1)^2) = -n$, so it can be absorbed in a redefined coupling constant in the same way as $\gamma$:

$$\lambda\mu^{-2\varepsilon} \rightarrow \lambda\tilde{\mu}^{-2\varepsilon}, \tilde{\mu} = \mu\exp(\varepsilon(\gamma+\epsilon\zeta(2)/2),$$

an argument which is familiar from $\overline{MS}$ schemes.

More subtle is the cancellation of higher transcendental $\zeta$'s. As an example let us consider $\zeta(n-1)$, the highest possible transcendental appearing in a $n$-loop calculation. Only the highest pole $\frac{1}{\varepsilon^n}$ can generate it, so we have to consider

$$\frac{1}{n!}\frac{1}{\varepsilon^n}\frac{\zeta(n-1)}{n-1}\varepsilon^{n-1}[n+1+(-n)^{n-1}-(n+1)^{n-1}],$$

which is the contribution of $P_n$ to $\zeta(n-1)$.

Also the counterterm expressions have to provide their highest pole. Let us study a 4-loop example:

$$\begin{aligned} Z_1^{(4)} &= <P_4-<P_1>P_3-<P_2-<P_1>P_1>P_2 \\ &\quad -<P_3-<P_1>P_2-<P_2-<P_1>P_1>P_1>P_1>. \qquad (64) \end{aligned}$$

It follows for the coefficient of $\zeta(3)$:

$$\begin{aligned} S_4(3) &:= \frac{1}{\varepsilon}\frac{1}{4!0!}\frac{1}{3}[5+(-4)^3-5^3] \\ &\quad -\frac{1}{\varepsilon}\frac{1}{3!1!}\frac{1}{3}[4+(-3)^3-4^3] \\ &\quad -\frac{1}{\varepsilon}\frac{1}{2!2!}\frac{1}{3}[3+(-2)^3-3^3] \\ &\quad -\frac{1}{\varepsilon}\frac{1}{1!1!2!}\frac{1}{3}[3+(-2)^3-3^3] \\ &\quad -\frac{1}{\varepsilon}\frac{1}{3!1!}\frac{1}{3}[2+(-1)^3-2^3], \\ &\quad \ldots \end{aligned}$$

where ... refers to the last three terms in Eq.(64) which add to zero. We have for the coefficient of $\zeta(3)$

$$S_4(3)=\frac{1}{\varepsilon}\sum_{i=1}^{4}(-1)^i\frac{1}{i!(4-i)!}\frac{1}{3}[i+1+(-i)^3-(i+1)^3]=0.$$

It is easy to see that in general the coefficient of $\zeta(n-1)$ is given by

$$S_n(n-1):=\frac{1}{\varepsilon}\frac{1}{n-1}\sum_{i=1}^{n}(-1)^i\frac{1}{i!(n-i)!}[i+1+(-i)^{n-1}-(i+1)^{n-1}].$$

So we have to show $S_n(n-1)=0$. Let us use

$$(1-a)^n=\sum_{i=0}^{n}\frac{(-1)^i a^i n!}{i!(n-i)!}=\sum_{i=0}^{\infty}\frac{(-1)^i a^i\Gamma(n+1)}{\Gamma(i+1)\Gamma(n-i+1)},$$

from which we conclude

$$\delta_{n,0}=\sum_{i=0}^{\infty}\frac{(-1)^i}{\Gamma(i+1)\Gamma(n-i+1)}.$$

Consider

$$T_n(r):=\sum_{i=0}^{\infty}\frac{(-1)^i i^r}{\Gamma(i+1)\Gamma(n-i+1)},\quad r>0.$$

which appears in $S_n(n-1)$. It is possible to express $i^r$ as a linear combination of terms

$$i(i-1)\ldots(i-r+1)+i(i-1)\ldots(i-r+2)+\ldots+i,$$

so that

$$i^r = \sum_{j=1}^{r} c_{rj} \frac{\Gamma(i+1)}{\Gamma(i-j+1)},$$

and we obtain

$$\begin{aligned}
T_n(r) &= \sum_{i,j=1}^{r} (-1)^i \frac{c_{rj}}{\Gamma(i-j+1)\Gamma(n-i)} \\
&= \sum_{i,j}^{r} (-1)^{i+j} \frac{c_{rj}}{\Gamma(i+1)\Gamma(n-i-j)} \\
&= \sum_{j=1}^{r} (-1)^j c_{rj} \delta_{n-j,0} \\
&= (-1)^n c_{rn}.
\end{aligned}$$

We have

$$c_{rn} = 0, \; r < n,$$
$$c_{rn} = 1, \; r = n,$$

and finally

$$T_n(r) = 0 \text{ for } 0 < r < n,$$
$$T_n(n) = (-1)^n.$$

By a similar argument we can show that

$$\begin{aligned}
U_n(r) &:= \sum_{i=0}^{\infty} \frac{(-1)^i (i+1)^r}{i!(n-i)!} = 0 \text{ for } 0 < r < n, \\
U_n(n) &= (-1)^n.
\end{aligned}$$

Combining everything we find

$$S_n(r) = 0, \; r \leq n,$$

which includes the desired result for $S_n(n-1)$. It is a pleasure to thank R. Delbourgo for helping out with this proof.

Some comments might be appropriate. We investigated the above case for $D = 4 - 2\varepsilon$ dimensions, where our function ${}_j\Delta$ would not correspond to a Green function. Nevertheless it is the basic example for all possible cases in a renormalizable theory. In the case of tensor integrals or dimensions other than four our function would only be modified by a polynomial of $\varepsilon$ multiplying it. This leaves the basic structure of the sums above unaffected and the reasoning remains unchanged. We could now proceed to show the absence of the other transcendentals $S_n(r)$ along similar lines. But this is a cumbersome technical task.

We rather like to give a more general argument establishing rational contributions for $Z$ factors from simple topologies for all renormalizable theories.

We proceed by induction. Assume that we have established our desired result at the $n$-loop level. At the $(n+1)$-loop level we are confronted with the $(n+1)$-loop graph and all its counterterm graphs, including the $n$-loop graph with its counterterms, which combine to purely rational divergences by assumption.

$$Z^{(n+1)} = < G^{(n+1)} - \sum_{i=1}^{n} Z^{(i)} G^{(i)} >, \tag{65}$$

with rational $Z^{(i)}$.

Note that for all $i$, $1 \leq i \leq n+1$, the $i$-th loop integration looks like

$$I^{(i)} = \int d^D k \frac{N(k)}{(k^2)^{(r+\varepsilon(i-1))}(k+q)^2}$$
$$= (q^2)^{D-r-1-\varepsilon i} \int d^D k \frac{N(k\ q)}{(k^2)^{(r+\varepsilon(i-1))}(k+\hat{e_q})^2};$$

$N(k\ q), r, T(k)$ are determined by the actual theory and Green function under consideration and are defined by the requirement that the case $i=1$ is the corresponding one-loop Green function. $N(k)$ takes into account a possible numerator structur and $N(k\ q)$ incorporates a scaling of the vector $k$ by the modulus of $q$. $T(q)$ is the corresponding tensor structur after integration.

Now we notice that we can render this expression finite in the following difference (higher order subtractions for divergences worse than logarithmic are understood)

$$\begin{aligned} I_\Delta^{(i)} &= I^{(i)} - (q^2)^{D-r-1-\varepsilon i} \int d^D k \frac{N(k)\Theta(\mid k^2 \mid -1)}{(k^2)^{(r+1+\varepsilon(i-1))}} \\ &=: I^{(i)} - I_{rat}^{(i)}, \end{aligned}$$

where we defined a new function $I_{rat}^{(i)}$. The integration of $I_{rat}^{(i)}$ gives (it is sufficient to study the case $N(k)=1$, extra numerator structures do not spoil the argument, but blow up the notation)

$$I_{rat}^{(i)} = (q^2)^{D-r-1-\varepsilon i} \frac{\pi^{(D/2)}}{\Gamma(D/2)} \frac{1}{D-2-2r-2\varepsilon i}.$$

The idea is to express the whole $Z$ factor at the $(n+1)$-loop level in terms of $I_{rat}$. The only source of transcendentality would then be $\pi^{D/2}/\Gamma(D/2)$, which is independent of the writhe number. So we redefine $I_{rat}^{(i)}$

$$I_{rat}^{(i)} \to J_{rat}^{(i)} = \frac{\Gamma(D/2)}{\pi^{D/2}} I_{rat}^{(i)},$$

which defines a function $J_{rat}^{(i)}$, rational to all orders in $\varepsilon$, which still renders $I^{(i)}$ finite:

$$I_\Delta^{(i)} := I^{(i)} - J_{rat}^{(i)} = \mathcal{O}(1).$$

We further note that we can always arrange a redefinition of the renormalization parameter $\mu \to \mu \exp(f(\varepsilon))$ such that $\mathbf{G}_R^{(n)}$, the renormalized Green function at the $n$th loop level, has a rational finite part. We can do so without changing the $Z$-factor at the $n$-loop level.

Now we replace the ultimate loop integration in $G^{(n+1)}$ by $I_\Delta^{(n+1)}$. In fact we do so not only in $G^{(n+1)}$ but also in all the counterterm contributions which remove subdivergences nested in $G^{(n+1)}$, that is in ultimate loop integrations in Green functions appearing in the sum Eq.(65). As usual we realize that we can replace the ultimate loop integration by $J_{rat}^{(i)}$ in Eq.(65).

Now consider $Z^{(n+1)}$: it consists of a term $Z^{(n)} J_{rat}^{(0)}$ which is rational by the assumption on the induction and by the construction of $J_{rat}^{(0)}$, and a term $\mathbf{G}_R^{(n)}$ to be concatenated with $J_{rat}^{(i)}$, $2 \leq i < n$, which is rational by our renormalization group argument.

Note that the above argument would and should not work for topologies which are not of the simple ladder type. Then necessarily at least one pair of rungs is crossed. Let us assume there are $m$ planar rungs before the non-planar crossing pops up. At some stage we have to calculate a counterterm graph where these $m$ first loop integrations are replaced by the countertem. Then the non-planar first loop integrations destroy the start of the above induction, that is after scaling out the exterior momentum even in the region of large overall momentum the angular integration still involved due to the non-planar topology generate transcendental terms.

The most general simple topology is a rainbow dressing of ladder topologies as we learned in previous sections. The above argument can be applied to this cases when we take into consideration that these dressings themselves generate rational $Z$-factors by the above argument and that the whole graph is still finite when using $I_\Delta$ for the skeleton graph and add the corresponding rational counterterms for the dressing.

So in this section we learned that the a very peculiar sort of topological simple Feynman diagrams will only generate rational contributions to $MS$ $Z$-factors. We are now in a position to investigate more general topologies, hopefully finding a correspondence between the topolgy of the Feynman graph, the transcendentals in its $Z$-factor contribution, and its associated knot.

# 7 Knots and Transcendentals

We have rational coefficients contributions for the divergences originating from the simple topologies. This is very good news, as we will show in a minute that non-simple topologies give us transcendental coefficients. These transcendentals often arise from the expansion of the $\Gamma$ function near unit argument (cf. Eq.(63)), so the most prominent transcendentals in $Z$-factors are Riemann $\zeta$ functions of odd integer argument [12], and references there.

The first non-trivial candidate would be $\zeta(3)$. Let us consider some three-loop graphs.

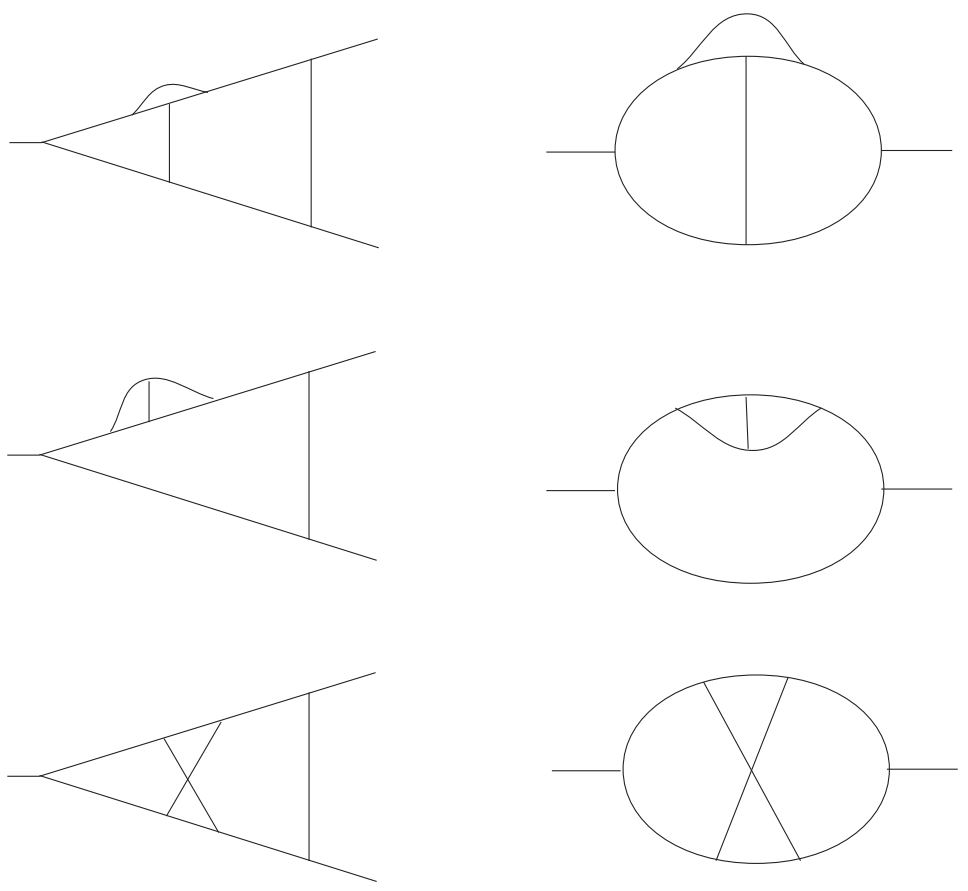

Fig.(38) These three-loop graphs involve $\zeta(3)$ in their divergent part.

It is well known that all the above graphs, even after adding their counterterms, give us a non-vanishing coefficient for $\zeta(3)$. This is fairly easy to see. For example the overlapping massless two-point function, a prominent example for generating a transcendental series in $\varepsilon$ [14], sitting as a subdivergence in several of the above graphs, gives as its finite value $6\zeta(3)$. The counterterm subtract only its divergent part, and so there remains the term proportional to $\zeta(3)$ multiplied by the divergence from the final loop integration.

The coefficient of $\zeta(3)$ naturally depends on the actual theory under consideration; we restrict ourselves to scalar theories at the moment for convenience. The corresponding link diagrams have a topology which for the first time involves a mutual entanglement of three links.

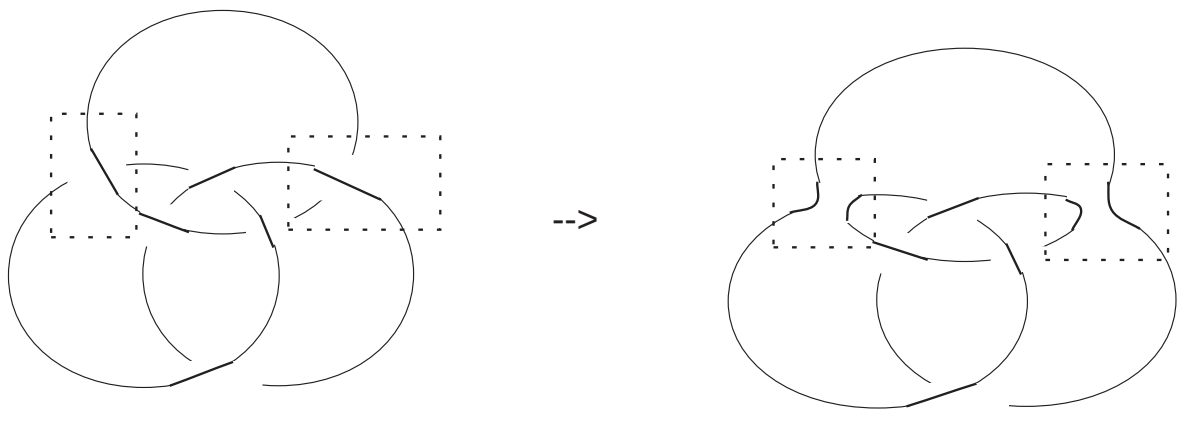

Fig.(39) The corresponding topology. The dashed rectangles indicates where the $Y$ part of the skein relation has been applied twice.

We see that the trefoil knot appears when we apply the $Y$ part of the skein relation twice. This topology gives us the braid expression

$$\sigma_1\sigma_2\sigma_1^2\sigma_2\sigma_1. \tag{66}$$

For the first time we encounter a more complicated word in braid group generators.

$$\begin{aligned}
&\sigma_1\sigma_2\sigma_1^2\sigma_2\sigma_1\\
&\rightarrow X\sigma_1\sigma_2\sigma_2\sigma_1+\sigma_1\sigma_2\sigma_1\sigma_2\sigma_1\\
&\rightarrow X^2\sigma_1\sigma_1+XY\sigma_1\sigma_2\sigma_1+YX\sigma_2\sigma_1^2+Y^2\sigma_1\sigma_2\sigma_1^2\\
&\rightarrow X^3\mathbf{1}+X^2Y\sigma_1+XYX\sigma_2+XY^2\sigma_1\sigma_2+YX^2\sigma_2\sigma_1\\
&+Y^2X\sigma_1\sigma_2+Y^3\sigma_1\sigma_2\sigma 1\\
&\rightarrow X^3\mathbf{1}+X^2Y\sigma_1+XYX\sigma_2+XY^2\sigma_1\sigma_2\\
&+YX^2\sigma_2\sigma_1+Y^2X\sigma_1\sigma_2+Y^3X\sigma_2+Y^4\sigma_1\sigma_2.
\end{aligned} \tag{67}$$

We had to use $\sigma_1\sigma_2\sigma_1=\sigma_2\sigma_1\sigma_2$. Note that in the third line above the expression $\sigma_1\sigma_2\sigma_1^2$ appears, corresponding to a trefoil knot with a twist.

Let us comment on the following two graphs.

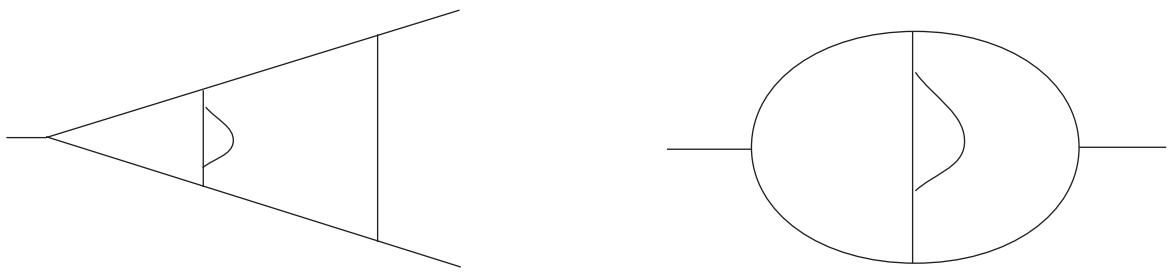

Fig.(40) Two examples which come near to a trefoil.

According to our rules this would also map to the topology in Fig.(39), so we would expect $\zeta(3)$ to be associated with the corresponding counterterm. But there is a difference between these graphs and the ones of Fig.(38). In Fig.(38) there is no way to translate loop momenta so that the topology becomes simple. In Fig.(39) we can easily do so, as long as not further dressings forbid this. So we expect $\zeta(3)$ only to be detected when these further dressings are present, an assertion which can be easily checked, investigating the function

$$G(r,s) = \int d^D k \frac{1}{k^{2-2r\varepsilon}(k+q)^{2-2s\varepsilon}}.$$

which differs from ${}_r\Delta$ in $\mathcal{O}(\varepsilon^3)$, as expected.

So we are prepared to assign $\zeta(3)$ to the trefoil topolgy. In the following we give some results of a detailed investigation of Feynman graphs calculated so far. Such a full classification will be given elsewhere [15], but we will report on some results from this work, which might serve as an introduction into the general scheme.

First, we note that we can assign loop momenta always in a way that they all encircle a given point $\bullet$ inside the diagram counterclockwise. This point corresponds to an axis in a closed braid diagram where all strands are oriented to encircle it counterclockwise. Having defined such an axis in the Feynman graph we replace the momentum flow by strands according to our rules and read off the braid group expression from the link diagram. For the trefoil example we find:

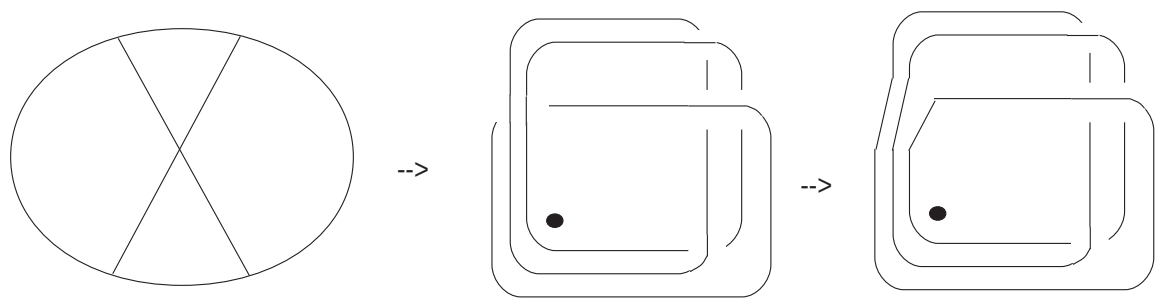

Fig.(41) An economic way from a Feynman diagram to a link diagram to a knot. We omit to indicate exterior momenta. So the Feynman graph on the left is generic to all the graphs of Fig.(37).

Note that the trefoil knot is the $(2,3)$ torus knot [9]. We can easily generalize this example to the $n$-loop case:

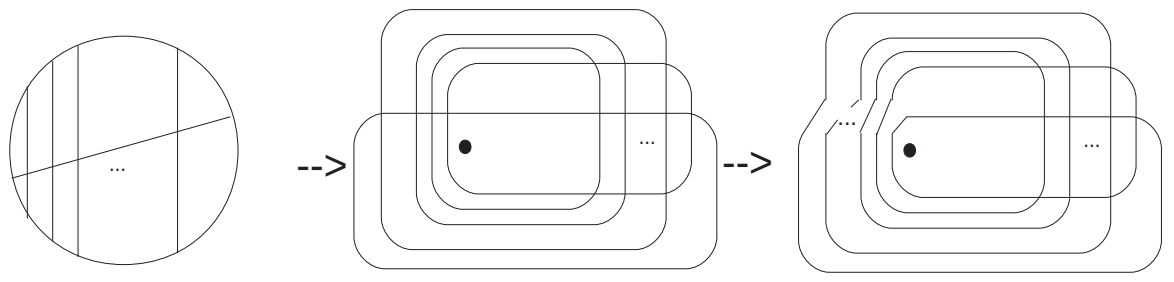

Fig.(42) The same for the $n$-loop case. Note that $n-1$ loops realize the ladder topology, and one loop goes through the middle of this $n-1$ cable.

It is a well known fact that these Feynman graphs are proprtional to $\zeta(2n-3)$ at the $n$ loop level [16]. The corresponding link diagram generates the knot $\sigma^{2n-3}$, which is the $(2,2n-3)$ torus knot. To see this we read off from the above picture the braid group expression:

$$\sigma_{n-1}\ldots\sigma_1\sigma_2\ldots\sigma_{n-1}\sigma_1\ldots\sigma_{n-2} = \sigma_{n-2}^{2n-3}$$

after applying Markov- and Reidemeister-moves. For example, chosing $n = 4$, we calculate

$$\begin{aligned}
\sigma_3\sigma_2\sigma_1\sigma_2\sigma_3\sigma_1\sigma_2 &= \\
\sigma_2\sigma_1\sigma_2\sigma_1\sigma_3\sigma_2\sigma_3 &= \\
\sigma_2\sigma_1\sigma_2\sigma_1\sigma_2\sigma_3\sigma_2 &= \\
\sigma_2\sigma_2\sigma_1\sigma_2\sigma_1\sigma_2\sigma_3 &= \\
\sigma_2\sigma_2\sigma_1\sigma_2\sigma_1\sigma_2 &= \\
\sigma_2\sigma_2\sigma_2\sigma_1\sigma_2\sigma_2 &= \\
\sigma_2^5\sigma_1 = \sigma_2^5&,
\end{aligned}$$

and one easily proves the result for arbitrary $n$ by induction [15].

So we have the beautiful correspondence $\zeta(2n-3)$ *in the Feynman graphs* $\leftrightarrow$ $(2, 2n-3)$ *torus knot in the link diagram.* Further correspondences generalizing such relations to more complicated topologies will be reported in [15].

Let us outline how to show that the correspondence between a ladder topology and the corresponding link diagram is unique. All ways to realize the ladder give the writhe number $n$ unknot for its connected component. But then, grouping all ladders in corresponding cables, it is easy to see that in general the assignment between Feynman graphs and link diagrams is unique, as we can directly read off the behaviour of a loop crossing an $r$-ladder: it groups together an $r$-link simple ladder topology and goes through this cable (which we can assume to live on a torus). We can group the result again to a cable and investigate the crossing of the next loop with this cable. A detailed analysis of this behaviour in terms of iterated tori links and its connection to the skeleton expansion will be given in future work.

Here we give two other typical results. First consider disjunct subdivergences:

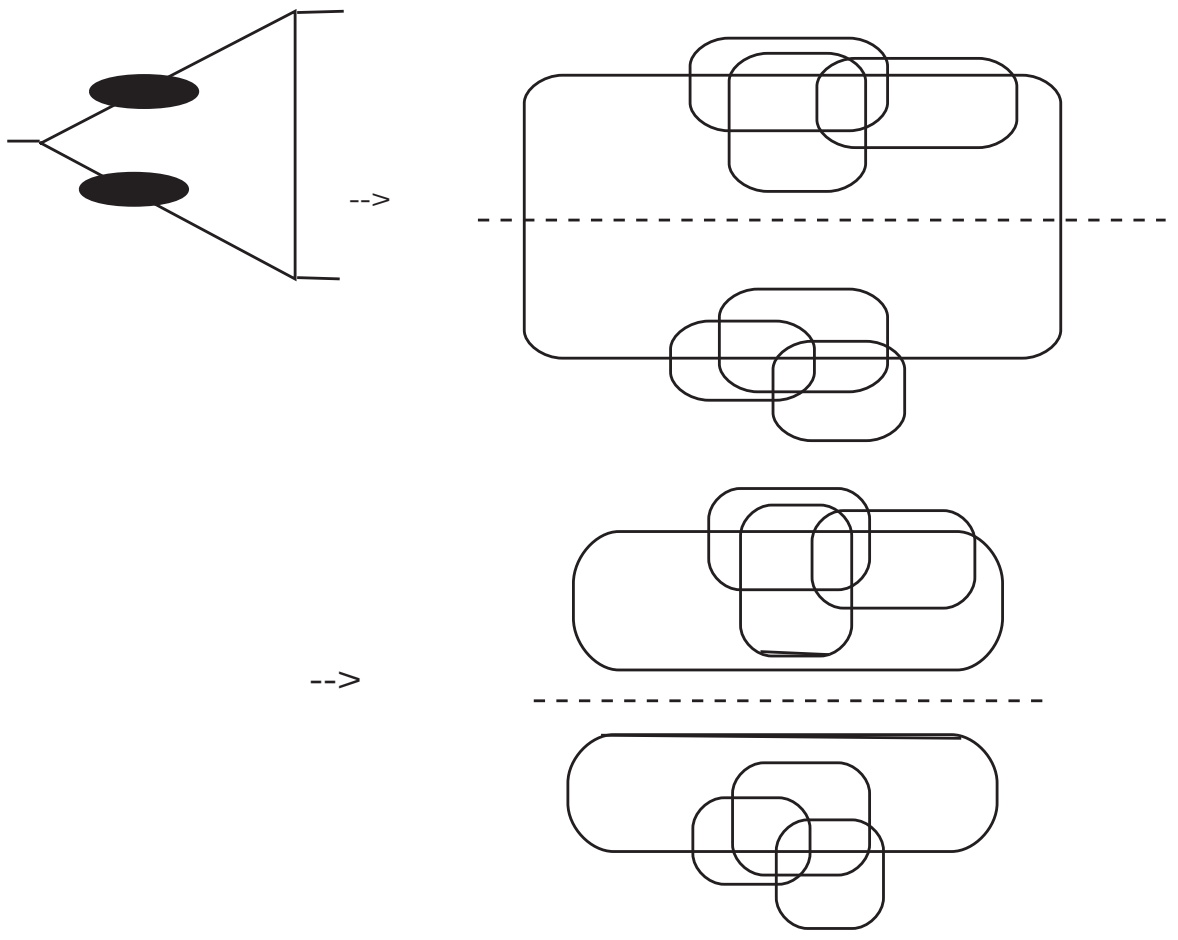

Fig.(43) These disjunct divergences produce factor knots. The dashed line only cuts two lines and separates two independent knots.

Disjunct divergences produce independent maximal forests. The above picture shows that this produces link diagrams which are 2-line-reducible. This is the defining condition for a factor knot, so that the corresponding transcendentals for both subdivergences correctly multiply.

Also relations between various graphs can be predicted:

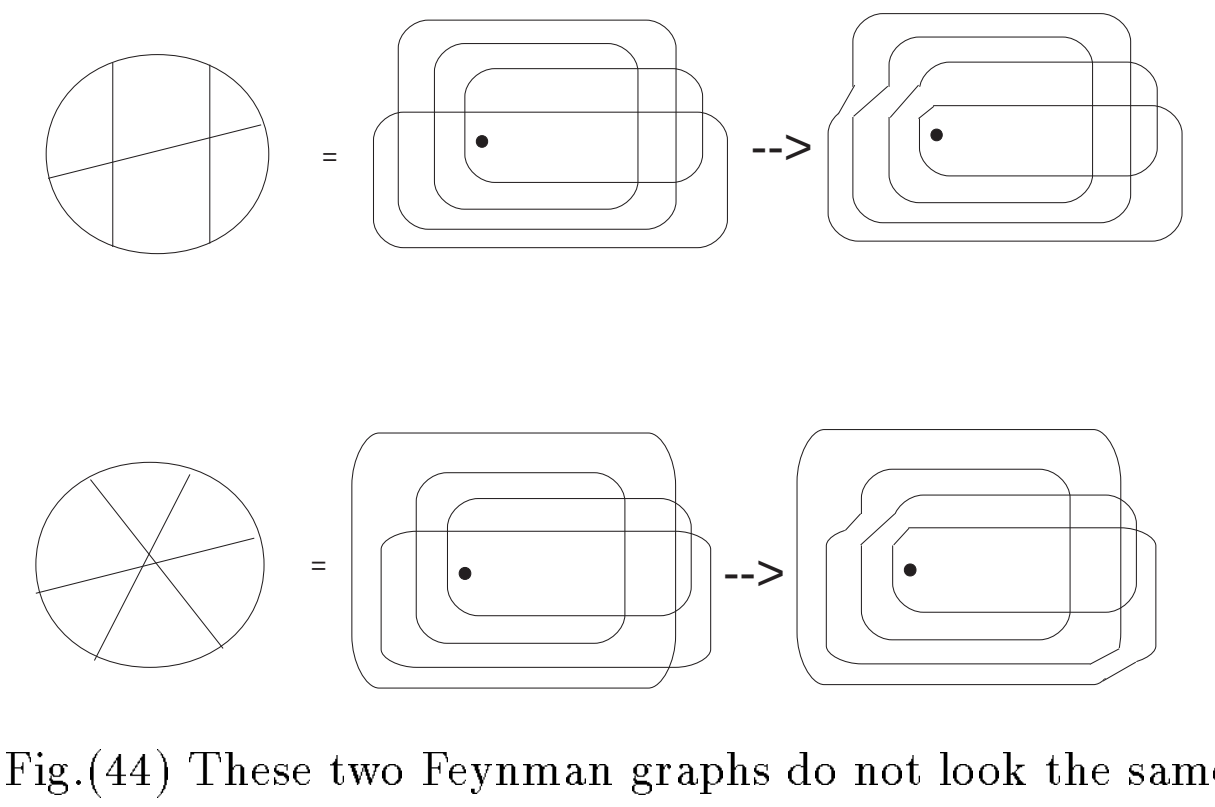

Fig.(44) These two Feynman graphs do not look the same, but both give $\zeta(5)$ [11]. It also follows immediately from the link diagram, by noticing that both generate the same knot.

So both give the (2,5) torus knot as reported above.

In this introductory paper we wanted to establish a connection between Feynman graphs and knot theory. Future work has to focus on a more systematic approach to the subject. We will comment now some more general features of our approach.

# 8 Comments, Conclusions and Summary

We have seen that topolgical information can be obtained from a Feynman graph. We found a way to map the Feynman graph to a link diagram, and associate a certain knot with this diagram. Results obtained so far indicate a one-to-one correspondence between the transcendental in the $Z$-factor contribution of this graph and the corresponding knot. The class of knots which appears is the class of iterated torus knots, a class of knots which plays a prominent role in algebraic topology. In future work we will not only try to incorporate powerful tools of braid algebra techniques into these considerations -that is for example looking for the $R$ matrix, but also investigate these connections to algebraic topology. Especially, we will investigate connections to topological field theories [17]. We hope that in the future it might be possible not only to identify the transcendentals in the $Z$-factor by considering the associated link diagram, but also hope to be able to predict the actuall coefficients of these transcendentals when the theory under investigation is specified. This would enable us to calculate $Z$-factors to arbitrary loop order just from knowledge of the topology of the graph and knowledge of its particle content. While this paper was written some progress along these lines has already been made and will be reported elsewhere.

Acknowledgements I am most grateful to Bob Delbourgo, Peter Jarvis and Ioannis Tsohantjis. Without their helpful hints and comments this paper would have never seen its final form. Most enjoyable discussions with David Broadhurst initiated the classification of Feynman diagrams referred to in previous sections.

This work was supported under grant A69231484 from the Australian Research Council.

# A Form Factors

The situation is most easily explained by stydying a example. Consider the case of a one-loop vertex correction in massless QED at zero momentum transfer:

$$\int d^D l \, \frac{1}{(l+k)^2}\gamma_\rho \frac{1}{\not l}\gamma_\mu \frac{1}{\not l}\gamma_\rho =$$
$$(\Delta_{11}\gamma_\mu + \Delta_{21}\not p p_\mu)(k^2)^{-\varepsilon},$$

which allows for two form factors $\Delta_{11}$ and $\Delta_{21}$. Only $\Delta_{11}$ is UV divergent.

Iterating this in nested topologies both form factors may replace the tree level vertex $\gamma_\mu$. So we also have to consider functions where the vertex structure is $\not{k} k_\mu$ for some momentum $k$, and this gives us two more $\Delta$ functions:

$$\int d^D l \, \frac{(l^2)^{-\varepsilon i}}{(l+k)^2} \gamma_\rho \frac{1}{\not{l}} \gamma_\mu \frac{1}{\not{l}} \gamma_\rho =$$
$$(k^2)^{-\varepsilon(i+1)} [{}_i\Delta_{11} \gamma_\mu + {}_i\Delta_{21} \frac{\not{k} k_\mu}{k^2}],$$
$$\int d^D l \, \frac{(l^2)^{-\varepsilon i}}{(l+k)^2} \gamma_\rho \frac{1}{\not{l}} \frac{\not{l} l_\mu}{l^2} \frac{1}{\not{l}} =$$
$$(k^2)^{-\varepsilon(i+1)} [{}_i\Delta_{12} \gamma_\mu + {}_i\Delta_{22} \frac{\not{k} k_\mu}{k^2}].$$

At the one-loop level we have

$$Z_1^{(1)} =< \Delta_{11} \gamma_\mu + \Delta_{21} \frac{\not{p} p_\mu}{p^2} >=< \Delta_{11} > \gamma_\mu,$$

by definition. At the next order we find

$$\begin{aligned} Z_1^{(2)} &=< (\Delta_{11} \; {}_1\Delta_{11} + \Delta_{21} \; {}_1\Delta_{12}) \gamma_\mu \\ &+ (\Delta_{11} \; {}_1\Delta_{21} + \Delta_{21} \; {}_1\Delta_{22}) \frac{\not{p} p_\mu}{p^2} > \\ &- << \Delta_{11} > (\Delta_{11} \gamma_\mu + \Delta_{21} \frac{\not{p} p_\mu}{p^2}) > \\ =< (\Delta_{11} \; {}_1\Delta_{11} + \Delta_{21} \; {}_1\Delta_{12}) \gamma_\mu > &- << \Delta_{11} > \Delta_{11} \gamma_\mu > \end{aligned}$$

as $< \Delta_{21} \Delta_{22} >= 0$. We used

$$< \Delta_{11} \; {}_1\Delta_{21} > - << \Delta_{11} > \Delta_{21} >= 0,$$

by Weinberg's theorem (we subtracted out the subdivergence from an overall convergent form factor $\sim \not{p} p_\mu$). UV-divergences only appear in the form factor $\sim \gamma_\mu$, as it should be. But we see that, at higher order, contributions of the other form factor mix into the result. Nevertheless, the concatenation works when we take this mixing iteratively into account. The above considerations are easily generalized to more complicated cases involving more different form factors, $r$ say. We would obtain an $r \times r$ matrix of $\Delta$ functions $\Delta_{ij}$, which concatenates in the way outlined above.

The same considerations apply to the two-point case, and the results of the example in the text concerning the two-loop fermion self-energy were obtained along these lines.

## B Cancellations of Transcendentals

We give an example for the concatenated product $B^6(\Delta)$, which corresponds to a seven loop calculation, compared with $(-A+B)^6(\Delta)$, for the example

$$\Delta = \int d^D k \, \frac{1}{k^2(k+q)^2} \,|_{q^2=1} \; .$$

We also give the lower order terms, at it is quite interesting to see the conspiration of rationals appearing. So in the following $G(r)$ is a $r+1$ loop Green function, and $Z(r)$ the corresponding counterterm expression. 'ge' is the Euler $\gamma$ and 'zet(i)' means $\zeta(i)$. We do not use a renormalization which would absorb the $\gamma$ and $\zeta(2)$, and also do not use that $\zeta(2)^3$, $\zeta(4)\zeta(2)$ and $\zeta(6)$ are all dependent ($\sim \pi^6$), as we want to exhibit the generated rationals in their purest form.

```
         1   -1    1   -2
Z(1):=---*x    - ---*x
         2         2

       -1            5      1   -2
G(1):=x  *( - ge + ---) + ---*x
                     2      2

         2   -1    1   -2    1   -3
Z(2):=---*x    - ---*x   + ---*x
         3         2         6

       -1       1            3    2    9        55
G(2):=x  *( - ---*zet(2) + ---*ge  - ---*ge + ----)
                4            4         2        6

          -2      1        3      1   -3
       + x  *( - ---*ge + ---) + ---*x
                  2        2      6

         5   -1   19   -2    1   -3    1    -4
Z(3):=---*x    - ----*x   + ---*x   - ----*x
         4         24         4         24

       -1      23              1                 7            4    3
G(3):=x  *( - ----*zet(3) + ---*zet(2)*ge - ---*zet(2) - ---*ge
                9              3                 6            9

             14    2    125        455
          + ----*ge  - -----*ge + -----)
              3          6         12
```

```
          -2      1             1    2    7        125
       + x  *( - ----*zet(2) + ---*ge  - ---*ge + -----)
                  12            3         3        24

          -3      1        7       1    -4
       + x  *( - ---*ge + ----) + ----*x
                  6        12      24

         14   -1    19   -2    11   -3    1    -4     1    -5
Z(4):=----*x    - ----*x   + ----*x   - ----*x   + -----*x
         5          12         24         12         120

       -1      133            335                335
G(4):=x  *( - -----*zet(4) + -----*zet(3)*ge - -----*zet(3)
               96             72                 18

               5         2    25          2    25
            + -----*zet(2)  - ----*zet(2)*ge  + ----*zet(2)*ge
               192            96                12

               245            125    4    125    3    1225    2
            - -----*zet(2) + -----*ge  - -----*ge  + ------*ge
               48             576         36          48

               595        6727      -2      67
            - -----*ge + ------) + x  *( - ----*zet(3)
                6          40               72

             5                5              25     3    25    2
          + ----*zet(2)*ge - ----*zet(2) - -----*ge  + ----*ge
             48               12             144         12

             245        119
          - -----*ge + -----)
             24          6

          -3      1             5     2    5        49
       + x  *( - ----*zet(2) + ----*ge  - ---*ge + ----)
                  48            48         6        24

          -4      1         1       1    -5
       + x  *( - ----*ge + ---) + -----*x
                  24        6      120

           -1    1313   -2    47   -3    25    -4    1    -5
Z(5):=7*x    - ------*x   + ----*x   - -----*x   + ----*x
                 360          48         144         48
```

```
             1    -6
         - -----*x
            720

       -1      512            183               1647
G(5):=x  *( - -----*zet(5) + -----*zet(4)*ge - ------*zet(4)
               75             80                160

               23                 23          2    207
            + ----*zet(3)*zet(2) - ----*zet(3)*ge  + -----*zet(3)*ge
               30                  5                 5

               4991           3         2       27          2
            - ------*zet(3) - ----*zet(2) *ge + -----*zet(2)
                45             80                160

               3            3    81           2    217
            + ----*zet(2)*ge  - ----*zet(2)*ge  + -----*zet(2)*ge
               20                40                20

               903             9     5    81    4    217    3
            - -----*zet(2) - -----*ge  + ----*ge  - -----*ge
               40             100         40         10

               2709    2    19369        62601      -2
            + ------*ge  - -------*ge + -------) + x  *(
                20            40           80

             61             23                69
          - -----*zet(4) + ----*zet(3)*ge - ----*zet(3)
             160            15                10

              1         2    3             2    27
          + -----*zet(2)  - ----*zet(2)*ge  + ----*zet(2)*ge
             160             40                40

             217            3     4     27    3    217    2    903
          - -----*zet(2) + ----*ge  - ----*ge  + -----*ge  - -----*ge
             120            40          20          20          20

             19369      -3      23             1
          + -------) + x  *( - ----*zet(3) + ----*zet(2)*ge
              240               90             40

             9             1     3    27    2    217         301
          - ----*zet(2) - ----*ge  + ----*ge  - -----*ge + -----)
```

```
             80            20         40         60         40

          -4       1             1     2    9         217
       + x  *( - -----*zet(2) + ----*ge  - ----*ge + -----)
                  240            40         40        360

          -5       1         3        1    -6
       + x  *( - -----*ge + ----) + -----*x
                  120        80      720

         132   -1    277   -2    839   -3    19   -4     7    -5
Z(6):=-----*x   - -----*x   + -----*x   - ----*x   + -----*x
          7          30          360          48         144

             1    -6      1     -7
         - -----*x   + ------*x
            240         5040

       -1      20641            49567               49567
G(6):=x  *( - -------*zet(6) + -------*zet(5)*ge - -------*zet(5)
               4320             3600                 720

               1687                  11809          2
            + ------*zet(4)*zet(2) - -------*zet(4)*ge
               5760                  5760

               11809              28679            102487        2
            + -------*zet(4)*ge - -------*zet(4) + --------*zet(3)
                576                480             12960

               5929                      5929
            - ------*zet(3)*zet(2)*ge + ------*zet(3)*zet(2)
               4320                      864

               41503          3    41503          2
            + -------*zet(3)*ge  - -------*zet(3)*ge
               12960                864

               100793             44891            49          3
            + --------*zet(3)*ge - -------*zet(3) - -------*zet(2)
                360                 72             34560

                343         2   2    343         2      833        2
            + -------*zet(2) *ge  - ------*zet(2) *ge + -----*zet(2)
               11520                1152                960

               2401           4    2401          3
```

```
  - -------*zet(2)*ge  + ------*zet(2)*ge
     34560                1728

     5831          2    2597              49049
  - ------*zet(2)*ge  + ------*zet(2)*ge - -------*zet(2)
     480                  48                  480

     16807     6    16807    5    40817    4    18179    3
  + --------*ge  - -------*ge  + -------*ge  - -------*ge
     518400         17280         2880          144

     343343    2    116039        4753177      -2
  + --------*ge  - ---------*ge + ---------) + x  *(
      480             48           1260

   7081           1687               1687
- ------*zet(5) + ------*zet(4)*ge - ------*zet(4)
   3600           2880               576

   847                   5929           2
+ ------*zet(3)*zet(2) - ------*zet(3)*ge
   4320                  4320

   5929               14399             49         2
+ ------*zet(3)*ge - -------*zet(3) - ------*zet(2) *ge
   432                 360             5760

    49         2    343             3    343           2
+ ------*zet(2)  + ------*zet(2)*ge  - -----*zet(2)*ge
   1152             8640                576

   833               371             2401     5    2401    4
+ -----*zet(2)*ge - -----*zet(2) - -------*ge  + ------*ge
   240               48              86400         3456

   5831    3    2597    2    49049        16577      -3
- ------*ge  + ------*ge  - -------*ge + -------) + x  *(
   720           48           240          48

   241            847                847
- ------*zet(4) + ------*zet(3)*ge - -----*zet(3)
   2880           2160               432

    7          2     49             2    49
+ ------*zet(2)  - ------*zet(2)*ge  + -----*zet(2)*ge
   5760             2880                288
```

```
     119             343     4    343    3    833    2
   - -----*zet(2) + -------*ge  - -----*ge  + -----*ge
     240            17280         864         240

     371        7007      -4      121
   - -----*ge + ------) + x  *( - ------*zet(3)
     24         240               2160

      7                  7              49    3    49     2
   + ------*zet(2)*ge - -----*zet(2) - ------*ge  + -----*ge
     1440                288            4320         288

     119        53
   - -----*ge + ----)
     120        24

    -5      1              7     2    7          17
+ x  *( - ------*zet(2) + ------*ge  - -----*ge + -----)
          1440            1440        144        120

    -6      1          1        1     -7
+ x  *( - -----*ge + -----) + ------*x
          720        144      5040
```